\documentclass[aps,twocolumn,10pt,aps,amsmath,amssymb,nofootinbib,superscriptaddress]{revtex4}
\usepackage{epsfig} 
\usepackage{amsmath} 
\usepackage{graphicx}
\usepackage{color}
\usepackage{float}
\usepackage{soul,xcolor}
\usepackage[font=scriptsize]{caption}
\usepackage{braket}
\usepackage{bbold}
\usepackage{subfig}
\usepackage{braket}
\usepackage{titlesec}
\usepackage{placeins}
\usepackage{hyperref}
\usepackage{caption}
\begin{document}
\title{Can NSI affect non-local correlations in neutrino oscillations?}

\author{Bhavna Yadav}
\email{yadav.18@iitj.ac.in}
\affiliation{Indian Institute of Technology Jodhpur, Jodhpur 342037, India}

\author{Trisha Sarkar}
\email{sarkar.2@iitj.ac.in}
\affiliation{Indian Institute of Technology Jodhpur, Jodhpur 342037, India}

\author{Khushboo Dixit }
\email{kdixit@iitb.ac.in}
\affiliation{Indian Institute of Technology Bombay, Mumbai 400076, India}

\author{Ashutosh Kumar Alok}
\email{akalok@iitj.ac.in}
\affiliation{Indian Institute of Technology Jodhpur, Jodhpur 342037, India}

\date{May 18, 2022}

\begin{abstract}
	Non-local correlations in entangled systems are usually captured by measures such as Bell's inequality violation. It was recently shown that in  neutrino systems, a measure of non-local advantage of quantum coherence (NAQC)  can be considered as a stronger measure of non-local correlations as compared to the Bell's inequality violation. In this work, we analyze the effects of  non standard interaction (NSI) on these measures in the context of two flavour neutrino oscillations for DUNE, MINOS, T2K, KamLAND, JUNO and Daya Bay experimental set-ups. We find that even in the presence of NSI, Bell’s inequality violation occurs in the entire energy range whereas the NAQC violation is observed only in some specific  energy range justifying the more elementary feature of NAQC. Further, we find that NSI can enhance the violation of NAQC and Bell's inequality parameter in the higher energy range of a given experimental set-up; these enhancements being maximal for the KamLAND experiment. However, the possible enhancement in the violation of the Bell's inequality parameter over the standard model prediction can be up to 11\% whereas for NAQC it is 7\%. Thus although NAQC is a comparatively stronger witness of nonclassicality, it shows lesser sensitivity to NSI effects in comparison to the Bell's inequality parameter.
	
\keywords{Quantum correlations \and Neutrino oscillations \and Nonstandard interactions}

\end{abstract}
\maketitle

\section{Introduction}

Quantum correlations which are based on the assumptions of locality and realism provide a platform to test and analyze the local hidden variable theories. The concept of quantum correlation was first discussed by Schrodinger \cite{schrodinger_1935}. Later, in 1964, John Bell mathematically formulated the concepts of locality and realism \cite{Bell:1964kc} and showed that if the classical picture is sufficient to describe the correlation existing between the outcomes of the measurement performed on two certain spatially separated systems, then they support Einstein's local hidden variable theory \cite{Einstein:1935rr} and in that case the correlation is required to obey a certain type of inequality, known as Bell-CHSH inequality \cite{Bell:1964kc}. Violation of this inequality indicates the non-local nature of correlations shared between two subsystems and therefore, gives an implication of its quantum nature \cite{PhysRevA.81.052318}. Thus, quantum correlations can be contemplated as excellent tools to identify whether a system acts as classically or quantum mechanically.

In a composite system, correlations can exist among its subsystems whether they are estranged by spatial or temporal separation. Some of the well known measures of spatial quantum correlations are based on entanglement-entropy \cite{Amico:2007ag,horodecki2009quantum}, quantum discord \cite{Ollivier:2001fdq}, steering \cite{paul2020shareability,uola2020quantum}, non-locality along with the well known Bell-CHSH inequality \cite{PhysRevLett.23.880}, while the temporal correlations encompasses Leggett-Garg inequality (LGI) \cite{Leggett} and Leggett-Garg type inequalities (LGtI) \cite{Emary}. A recent measure of nonclassicality can be considered to be the potential to achieve \textit{non-local advantage of quantum coherence} (NAQC) which exhibits the ability to steer the quantum coherence of a bipartite state in the presence of entanglement between two subsystems \cite{PhysRevA.95.010301}. The advantage of NAQC over Bell inequality is that for the systems to achieve NAQC they must violate Bell inequality \cite{Hu, Ding, Ming:2020nyc} which indicates NAQC as a stronger measure of non-local correlation. Previously various measures of nonclassicality have been studied in different optical and electronic systems \cite{PhysRevLett.47.460, PhysRevA.57.3229, PhysRevLett.111.100504} as well as in high energy particle systems such as neutral mesons and neutrinos \cite{Caban:2006ij,bramon2007review,Hiesmayr:2007he,PhysRevD.77.096002,blasone2009entanglement,Blasone:2014jea,Blasone:2015lya,Gangopadhyay:2013aha,nikitin2015proposal,ALOK201665,banerjee2015quantum,banerjee2016quantum,Formaggio:2016cuh,Fu:2017hky,Naikoo:2017fos,Cervera-Lierta:2017tdt,Kerbikov:2017spv,Naikoo:2018vug,Richter-Laskowska:2018ikv,Song:2018bma,Wang:2020vdm,Ettefaghi:2020otb,Ming:2020nyc,Li:2021fft,Blasone:2021cau,Blasone:2021mbc,Shafaq:2021lju}. In the scenario of neutrino oscillation, one can map the flavour state of neutrino-system as a single-particle mode-entangled state where distinct flavours of neutrino serve as different modes of oscillations and entanglement can be seen among these flavour-modes \cite{blasone2009entanglement,Blasone:2014jea}. Moreover, various nonclassical features can be expressed in terms of neutrino survival and oscillation probabilities \cite{dixit2018quantum,ALOK201665,Song:2018bma,Ming:2020nyc} that proclaims these measures as legitimate quantities to be studied for neutrino-systems.

Most of these studies detailed to quantum correlations in neutrino oscillations assume Standard Model (SM) interactions. Recently, the $l_1$-norm based coherence measure that can be considered as a cornerstone for all the other nonclassical correlations, is shown to be sensitive to possible new physics effects in the neutrino system \cite{Dixit:2019swl} that provides motivation for various quantum correlation measures to be reassessed in the presence of NSI effects. In this direction, couple of analyses in the presence of NSI effects are obtained considering temporal correlations in terms of LGtI in the context of neutrino oscillations \cite{Sarkar:2021qrz,poonam}.

In this work, we analyse the effects of NSI on spatial correlations exhibited by the neutrino-system employing Bell's inequality and NAQC measures in two flavour neutrino oscillation scenario in the context of various accelerator experiments  such as DUNE, MINOS, T2K; and reactor experimental set-ups viz. KamLAND, JUNO and Daya Bay. A complete oscillation framework should include three flavours of neutrino. However, in the context of different experimental configurations three flavour structure can be safely reduced to two flavour neutrino oscillation scheme. Moreover, these measures of quantum correlations are defined for a qubit set-up which can only be realized in the context of two flavour neutrino oscillations. 

The NSI can exhibit its effect at the sub-leading level in neutrino oscillation phenomena and hence can be visible in several current and upcoming neutrino experimental data. The SM Lagrangian consists of operators of canonical dimension $d \leq 4$. The  NSI effects can be included  in a model independent way within the framework of effective field theory by adding the non-renormalizable terms containing higher dimensional operators ($d > 4$) to the SM effective Lagrangian. In this work we have considered the effects of dimension-6 four-fermion operator.

The plan of this work is as follows. In section \ref{sec2} we briefly describe two nonclassicality witnesses in terms of Bell's inequality and NAQC. Then we discuss the formalism of our work in which we depict the mode-entanglement in two-flavour neutrino oscillation scheme in section \ref{sec3a}. The dynamics of the neutrino oscillation in standard matter and in the presence of NSI is provided in sections \ref{sec3b} and \ref{sec3c}, respectively. Then in section \ref{sec4} we discuss the results and finally infer our conclusions in section \ref{sec5}.

\section{Quantum Correlation Measures} \label{sec2}
In this section, we briefly discuss some measures of quantum correlations incorporated in this work.

\emph{Bell's Inequality}: It examines the quantum correlation existing between measurement-outcomes performed on two spatially separated counterparts of a composite system. Violation of this inequality exhibits non-local correlations embedded in the system which is one of the most crucial aspects of quantum theory. Maximal violation is realized by the largest eigenvalue of a Hermitian operator, known as the Bell operator 
\cite{PhysRevLett.68.3259}. A standard form of Bell-inequality is CHSH (Clauser-Horne-Shimony-Holt) inequality which is expressed as \cite{PhysRevLett.23.880}
\begin{equation}\label{bell1}
-2\leq \mathcal{B}_{CHSH} \leq 2.
\end{equation}

Here $\mathcal{B}_{CHSH}$ is known as the Bell operator. For a system consisting of two spin-1/2 particles A and B, 
the combined state with Hilbert space defined as $\mathcal{H}=\mathcal{H}_A\otimes\mathcal{H}_B$, is expressed in terms of the density matrix ($\rho$) as follows \cite{1995PhLA..200..340H}
\begin{equation}\label{bell2}
	\small
\rho=\frac{1}{4} \left [ I\otimes I+(r.\sigma )\otimes I+I\otimes(s.\sigma )+\sum_{A,B=1}^{3}T_{AB}(\sigma _{A}\otimes \sigma _{B}) \right].
\end{equation}
Here $I$ is the $2\times2$ identity matrix, $r$, $s$ are the vectors in real space and $\sigma$ stands for Pauli matrices. $T$ is the correlation matrix and its elements are given by $T_{AB}=Tr[\rho (\sigma _{A}\otimes \sigma _{B})]$. A real matrix can be constructed as $T^{\dagger}T$ having eigenvalues $u_i$ ($i=1,2,3$) of which two largest positive eigenvalues are taken into account, denoted by $u_i$ and $u_j$. Violation of Bell-CHSH inequality given in Eq. \eqref{bell1} is possible {\it iff} $M(\rho) = u_i + u_j > 1$ \cite{1995PhLA..200..340H}. In the context of two flavour neutrino system $M(\rho)$ can be expressed as a simple algebraic function of the survival ($P_{surv}$) and oscillation probabilities ($P_{osc}$) as follows \cite{ALOK201665}
\begin{equation}\label{bell3}
M(\rho)=1+4P_{surv}P_{osc}.    
\end{equation}
It is noticeable that the maximal violation of $M(\rho)$ will be observed for $P_{surv} = P_{osc} = 1/2$.

\par
\emph{Non-local Advantage of Quantum Coherence}: 
Another essential attribute of a quantum system is the presence of coherence. Recently, the concept of non-local advantage of quantum coherence \textit{i.e.} NAQC was introduced in \cite{PhysRevA.95.010301, 2020QuIP...19..375X} that can provide an indication of the existence of quantumness in terms of entanglement and steerability in the system.  
Coherence of a system represented by the state $\rho$ can be quantified by $l_1$ norm which in the eigen basis of Pauli spin matrix $\sigma_i$ ($i=x,y,z$) is defined as \cite{PhysRevA.95.010301}
\begin{equation}
C_{l_1}^i(\rho)=\sum\limits_{i_1,i_2}|\bra{i_1}\rho\ket{i_2}|, (i_1\neq i_2).
\end{equation}\label{naqc1}
Here $\ket{i_1}$ and $\ket{i_2}$ are the eigen vectors of $\sigma_i$. Then the upper limit of the following quantity is given by \cite{PhysRevA.95.010301}
\begin{equation}\label{naqc2}
\sum\limits_{i=x,y,z} C_{l_1}^i(\rho) \leq \sqrt{6} \approx 2.45.
\end{equation}
In Eq. \eqref{naqc2}, the equality holds for the pure state which is a superposition of all the mutually orthonormal states with equal coefficients \emph{e.g.} $\rho=\frac{1}{2}\left[I+\frac{1}{\sqrt{3}}(\sigma_x+\sigma_y+\sigma_z)\right]$, where $I$ is the $2\times2$ identity matrix. For a single system description, the above equality holds.  
\par
To perceive the concept of NAQC, let us consider an entangled state, consisting of two subsystems $A$ and $B$, expressed by the density matrix $\rho$ as given in Eq. \eqref{bell2}. If one performs a random measurement on the subsystem $A$ in the eigen basis of $\sigma_i$ ($i=x,y,z$), then the probability of the outcome, $a\in \{0,1 \}$ is given by, $p(\rho_{B|\Pi_i^a})=$Tr$[(\Pi_i^a \otimes I)\rho]$ where $\Pi_i^a=[I+(-1)^a \sigma_i]/2$ and $\rho_{B|\Pi_i^a}$ is the conditional state of $B$ defined as, $\rho_{B|\Pi_i^a}=~Tr_A(\rho_{|\Pi_{i}^{a}})$ which is obtained after tracing out the effect of $A$. Such measurement on $A$ affects the coherence of the subsystem $B$. Another measurement may be performed on the subsystem $B$ at random in the eigen basis of other two Pauli matrices (say, $\sigma_j$, $\sigma_k$; $j,k \neq i$).  
The violation of Eq. \eqref{naqc2} infers the fact that the single system description of the coherence of the subsystem $B$ is not feasible. Therefore NAQC of the state $B$ is achieved by the condition
\begin{equation}\label{naqc3}
N_{l_1}(\rho)=\frac{1}{2}\sum\limits_{i,j,a}p(\rho_{B|\Pi_i^a})C_{{l_1}}^i(\rho_{B|\Pi_i^a})>\sqrt{6}.
\end{equation}
Any state defined by $\rho$ may not necessarily achieve such non local advantage of quantum coherence even if it is entangled. Canonically NAQC is the stronger measure compared to Bell's inequality. In case of two flavour neutrino system, the two subsystems, $A$ and $B$ can be realized by the two flavours of the neutrino. To analyze the two nonlocal measures in context of two flavour neutrino  system, we discuss about the mode entanglement in the next section.

\section{Formalism} \label{sec3}
In this section, we present the theoretical framework
of our analysis. We start with a brief description of mode entanglement in section \ref{sec3a}. Then we discuss the dynamics of neutrino
oscillations under the effect of both SM interaction  and NSI in section \ref{sec3b} and \ref{sec3c}, respectively. 

\subsection{Mode Entanglement} \label{sec3a}
 
It is a  well known fact that neutrino oscillation requires the flavour eigenstates $\nu_{\alpha}$ to be represented as a linear combination of mass eigenstates $\nu_{i}$ as follows
\begin{equation}\label{1}
    \ket{\nu_{\alpha}}= \sum_{i}U_{\alpha i}\ket{\nu _{i}}, 
\end{equation} 
where $U$ represents the elements of the unitary mixing matrix. The time evolution of mass eigenstates is given by
\begin{equation}\label{2}
    \ket{\nu_{i}(t)}=e^{-\iota  E_{i}t}\ket{\nu_{i }},
\end{equation}
where $\ket{\nu_{i}}$ are mass eigenstate at $t=0$ given by Eq. \eqref{1}. Therefore the time evolved flavour eigen state is expressed as
\begin{eqnarray}\label{3}
    \ket{\nu _{\alpha}(t)}=\sum_{i}  U_{\alpha i} e^{-\iota  E_{i}t} \ket{\nu_{i}} \notag \\
=\sum_{i,\beta}  U_{\alpha i} e^{-iE_{i}t} U_{\beta i}^{*} \ket{\nu_{\beta}}.
\end{eqnarray}
$\ket{\nu_{\alpha}}$ and $\ket{\nu_{i}}$ denote the flavour and mass eigen states respectively at $t=0$. In the relativistic limit, neutrino flavour states are considered to be individual modes. In the two flavour neutrino system, it can be expressed as \cite{blasone2009entanglement}
\begin{equation}\label{m4}
    \ket{\nu _{\alpha}}\equiv \ket{1}_{\alpha} \ket{0}_{\beta }\equiv\ket{10}_{\alpha \beta},\hspace{0.35 cm}\ket{\nu}_{\beta }\equiv\ket{0}_{\alpha } \ket{1}_{\beta }\equiv\ket{01}_{\alpha \beta}.  
\end{equation} 
Here $\ket{m}_{\gamma}$ gives the occupation number of a particular state in $\gamma$-th mode ($m=\{0,1\}~,\gamma=\{\alpha,\beta\}$). For example, $\ket{1}_{\alpha}\ket{0}_{\beta}$ implies the presence of the neutrino in $\alpha$-th mode and its absence in $\beta$-th mode. Such representation explicitly shows the entanglement of different modes or flavours in a single particle state. Using Eq. \eqref{m4} in Eq. \eqref{3}, we get
\begin{equation}\label{5}
    \ket{\nu _{\alpha }(t)}=\Bar{U}_{\alpha \alpha}(t) \ket{1}_{\alpha} \ket{0}_{\beta}+\Bar{U}_{\alpha \beta}(t) \ket{0}_{\alpha} \ket{1}_{\beta},
\end{equation}
where $\Bar{U}=U e^{-i \mathcal{H}_m t} U^{\dagger}$, $\mathcal{H}_m = diag(E_1,E_2)$ {\it i.e.,} the Hamiltonian in mass eigenbasis with eigen values $E_{i}$ $(i=1,2)$. Eq. \eqref{5} indicates the entanglement between flavour-modes in a single particle system at time $t$. Then, the density matrix corresponding to the state given in Eq. \eqref{5} is expressed as
\begin{equation}\label{6}
\rho(t)=\begin{pmatrix}
0 & 0  & 0  & 0 \\ 
0 &\left | \Bar{U}_{\alpha \alpha }(t) \right |^{2}  & \Bar{U}_{\alpha \alpha }(t)\Bar{U}^{*}_{\alpha \beta }(t) &0 \\ 
0 & \Bar{U}^{*}_{\alpha \alpha }(t)\Bar{U}_{\alpha \beta }(t) & \left | \Bar{U}_{\alpha \beta }(t) \right |^{2} &0 \\ 
0 & 0 & 0 & 0
\end{pmatrix},
\end{equation}
In the notion of two flavour neutrino oscillation, the $2\times 2$ mixing matrix $U$ is represented as
\begin{equation}\label{7}
U=\begin{pmatrix}
U_{\alpha 1} & U_{\alpha 2} \\ 
U_{\beta 1} & U_{\beta 2} 
\end{pmatrix}
=\begin{pmatrix}
 \cos \theta & \sin \theta \\ 
 - \sin \theta & \cos \theta
\end{pmatrix},
\end{equation}
where $\theta$ is the mixing angle. The coefficients in Eq. \eqref{5}, $\Bar{U}_{\alpha \alpha}$ and $\Bar{U}_{\alpha \beta}$ can be expressed as
\begin{eqnarray}
\Bar{U}_{\alpha\alpha} = \cos^{2} \theta \, e^{-iE_{1}t} + \sin^{2} \theta \, e^{-iE_{2}t}, ~~~~~~~~~~~~\\
\Bar{U}_{\alpha\beta} = -\sin \theta \cos \theta \, e^{-iE_{1}t} + \sin \theta \cos \theta \, e^{-iE_{2}t}.
\end{eqnarray}

In ultra relativistic limit, ~where $L\equiv t$, $L$ and $E$ being the distance traversed by the neutrino and neutrino-energy, respectively and $E_i -E _j \approx \Delta m_{ij}^{2}/2 E$, provided that the two mass eiegnstates travel with equal momenta ($E=\Vec{p_1}=\Vec{p_2}$).

\subsection{Neutrino Oscillation in Matter} \label{sec3b}

While travelling through the matter neutrinos undergo charged current (CC) and neutral current (NC) interactions with matter particles. This can affect the flavour oscillation significantly. Neutrinos can undergo both coherent and incoherent scattering inside a medium, however, the effect of incoherent scattering is neglected because of the large mean free path corresponding to it. Also since the Earth matter is composed of only nucleons and electrons, the contribution to CC interaction comes only from electron, while NC interaction involves only the neutrons as the NC potential of electron and proton are cancelled due to the assumption of charged neutrality of matter. The effect of NC can be neglected since it provides equal contribution for each flavour and hence generates an overall phase factor that does not affect the transition probability. For an incoming $\nu_e$ traversing through Earth, the corresponding Hamiltonian in mass basis is given by \cite{Giunti:2007ry}
\begin{equation}\label{osc3}
\mathcal{H}_m=\mathcal{H}_{vac}+\mathcal{H}_{mat}=\begin{pmatrix}
E_{1} & 0 \\ 
0 & E_{2}
\end{pmatrix}+
U^\dagger\begin{pmatrix}
A & 0 \\ 
0 & 0 
\end{pmatrix}U,
\end{equation}
where $A=\pm \sqrt{2}G_{F}N_{e}$ is the standard matter potential. $G_{F}$ and $N_e$ are the Fermi constant and electron number density in matter, respectively. $A$ is positive for neutrinos and negative for anti-neutrinos. $\mathcal{H}_{vac}$ is the counterpart corresponding to the vacuum oscillation. If the oscillation scenario involves $\nu_\mu$ and $\nu_\tau$ only, the second term ($\mathcal{H}_{mat}$) in the R.H.S of Eq. \eqref{osc3} will vanish. The time evolution operator for neutrino mass eigen state is exhibited by $U_{m}(L) = e^{-i \mathcal{H}_mL}$ in the ultra-relativistic limit. 
 In the case of two flavour neutrino oscillations,  the evolution operator in the mass eigen basis is represented as \cite{Ohlsson:1999xb}
\begin{eqnarray}\label{osc4}
U_{m}(L)&=&e^{-i\mathcal{H}_mL}=\phi ~e^{-iLT}\nonumber\\
&=&\phi \sum_{a=1}^{2} e^{-i L\lambda _{a}}\frac{1}{2\lambda _{a}}\left ( \lambda _{a} I+T\right ).
\end{eqnarray}
Here $\lambda _{1}$ and $\lambda _{2}$ are the eigenvalues of the matrix $T$ ($\lambda_1=-\lambda_2$), which is expressed as
\begin{equation}\label{osc5}
T \equiv \mathcal{H}_{m}- (tr H_{m})I/2 =
\begin{pmatrix}
T_{11} & T_{12} \\
T_{21} & T_{22} 
\end{pmatrix},
\end{equation}
where
\begin{eqnarray}
T_{11} &=& A \cos^2\theta+\frac{1}{2}(E_1-E_2-A), \nonumber\\
T_{12}&=&T_{21}= A \cos\theta \sin\theta, \nonumber\\
T_{22}&=& A \sin^2\theta-\frac{1}{2}(E_1 - E_2 + A)\,.
\end{eqnarray}

The flavour evolution operator can be calculated as $U_{f}(L)=U^{\dagger} U_{m}(L) U$, where $U$ is the mixing matrix given in Eq. \eqref{7}. In our analyses, the Earth matter density is considered as $\rho= 2.8$ gm/cc which corresponds to the density potential $A \approx 1.01 \times 10^{-13}$ eV.

\subsection{Non Standard Interaction in neutrino oscillation} \label{sec3c}
NSI comprises the effect of new physics beyond the SM which plays the role of subleading effects for neutrino flavour oscillations \cite{Wolfenstein,Mikheev}. As we are entering the precision era, such subleading effects can be estimated with  higher accuracy. NSI can occur in both CC and NC interaction channels
and is expressed by the following four fermion dimension-6 operator \cite{Farzan:2017xzy,Davidson:2003ha,Antusch:2008tz,Dev:2019anc,Ohlsson:2012kf,Babu:2019mfe}
\begin{eqnarray}\label{nsi1}
\mathcal{L}_{NSI}^{CC}=2\sqrt{2}G_F\sum\limits_{\alpha, \beta, P} \epsilon_{\alpha\beta}^{ff^{'},P}(\bar{\nu}_\alpha \gamma^{\mu} P l_\beta)(\bar{f^{'}} \gamma_\mu P f),\nonumber \\ 
\mathcal{L}_{NSI}^{NC}=2\sqrt{2}G_F\sum\limits_{\alpha, \beta, P} \epsilon_{\alpha\beta}^{f,P}(\bar{\nu}_\alpha \gamma^{\mu} P \nu_\beta)(\bar{f} \gamma_\mu P f).~~~~
\end{eqnarray}
Here $P \in \{P_{L},P_R\}$, $P_{L,R}=(1\mp \gamma^5)/2 $. $P_L$ and $P_R$ are the left and right handed chirality operators, respectively.
$\epsilon_{\alpha \beta }^{f{f}'}$ is the dimensionless coefficient which measures the strength of NSI compared to weak interaction coupling constant $G_F$ $i.e.$ $\epsilon_{\alpha\beta}^{ff^{'},P}\sim \mathcal{O}(G_x/G_F)$. $\alpha$ and $\beta$ correspond to different neutrino flavours, $l_{\beta} = e, \mu, \tau$ and $\{f,f^{'} \}\in \{u,d \}$. 

In presence of NSI, the Hamiltonian given in Eq. \eqref{osc3} is modified as follows
\begin{eqnarray}\label{nsi2}
&\mathcal{H}_{tot}&=
\mathcal{H}_{vac}+\mathcal{H}_{mat}+\mathcal{H}_{NSI}
\nonumber\\&&=
\begin{pmatrix}
E_{1} & 0 \\ 
0 & E_{2}
\end{pmatrix}+U^{\dagger } A\begin{pmatrix}
b+\epsilon_{\alpha\alpha}(x) &\epsilon_{\alpha\beta}(x) \\ 
 \epsilon_{\beta\alpha}(x) & \epsilon_{\beta\beta}(x)
\end{pmatrix} U.~~~~~~~~
\end{eqnarray}

Here $b \in \{0,1\}$ depending on whether the transition includes $\nu_e$ state or not. $b=1$ if transition involves $\ket{\nu_e}$ state and $b=0$ otherwise.
$\epsilon_{\alpha\beta}(x)$ are the NSI parameters which are expressed as
\begin{equation}\label{nsi3}
    \epsilon _{\alpha \beta }(x)=\sum_{f=e,u,d}\frac{N_{f} (x) }{N_{e}(x)} \epsilon _{\alpha \beta }^{f}.
\end{equation}
Here $ N_{f} (x) $ is the fermion number density in matter and $x$ is the distance traversed by the neutrino. Taking charge neutrality condition into account ($N_p=N_e$), from the quark structure of neutron and proton we obtain, $N_{u}(x) = 2 N_{p}(x) + N_{n}(x)$ and $N_{d}(x) =  N_{p}(x) + 2 N_{n}(x)$. Using these conditions in Eq. \eqref{nsi3}, the expression for $\epsilon_{\alpha\beta}(x)$ is obtained to be
\begin{equation}
    \epsilon_{\alpha \beta }(x)=\epsilon_{\alpha \beta }^{e}+(2+Y_{n}(x))\epsilon_{\alpha \beta }^{u}+(1+ 2 Y_{n}(x))\epsilon_{\alpha \beta }^{d},
\end{equation}
where $Y_{n}=N_n(x)/N_e(x)$. NSI parameters are generally complex for which the flavour non-diagonal elements are not equal, while in case of real NSI, $\epsilon_{\alpha\beta}=\epsilon_{\beta\alpha}$. NSI can be both vector ($V$) and axial vector ($A$) type, $\epsilon_{\alpha\beta}^f=\epsilon_{\alpha\beta}^{f,L} \pm \epsilon_{\alpha\beta}^{f,R}$ ('$+$': vector, '$-$': axial vector). However only the vector counterpart is consistent with the neutrino oscillation in matter.
The bounds on NSI parameters are extracted from global analysis of the data obtained from different oscillation and non-oscillation experiments \cite{Esteban:2018ppq,Esteban:2019lfo,Coloma:2019mbs}. While CC NSI is strictly constrained, the bounds on NC NSI are comparatively weaker \cite{Biggio:2009nt}.

\section{Results and Discussion}\label{sec4}

In this section, we study the effects of NSI on Bell-CHSH parameter $M(\rho)$ and NAQC parameter $N_{l_1}(\rho)$  in the context of six experimental set-ups: (a) DUNE ($E\approx1-14$  GeV, $L \approx1300$ km) \cite{DUNE:2015lol,DUNE:2020jqi}, (b) MINOS ($E\approx1-10$ GeV, $L=735$ km) \cite{MINOS:2006foh}, (c) T2K ($E\approx0-6$ GeV, $L=295$ km) \cite{T2K:2011qtm,T2K:2013bzi}, (d) KamLAND ($E\approx1-16$ MeV, $L \approx180$ km) \cite{KamLAND:2002uet}, (e) JUNO ($E\approx1-8$ MeV, $L \approx53$ km) \cite{JUNO:2015zny,JUNO:2021vlw} and (f) Daya Bay ($E\approx0.8-6$ MeV, $L \approx2$ km)  \cite{DayaBay:2012fng,Roskovec:2020rgr}. Different experimental set-ups are sensitive to distinct parameter-spaces driving the two flavour-oscillations between $\nu_e-\nu_{\mu}$, $\nu_e-\nu_{\tau}$ or $\nu_{\mu}-\nu_{\tau}$. On the basis of sensitivity to neutrino oscillations in terms of different oscillation channels, we can classify experimental set-ups in three categories as follows

$(1)$ DUNE, MINOS and T2K are the long baseline (LBL) accelerator experiments having baselines equal to few hundreds of kilometers. DUNE and MINOS avail the NuMI $\nu_{\mu}$-beam situated at Fermilab as the source, while T2K is having Japan based J-PARC $\nu_{\mu}$-beam facility. These experiments are mainly sensitive to 2-3 sector of parameters, {\it i.e.,} $\theta_{23}$ and $\Delta_{32}$ (for appropriate approximations viz. \{$\theta_{12}$, $\Delta m_{21}^2\} \rightarrow 0$), driving the oscillation channel $\nu_{\mu}\rightarrow \nu_{\tau}$. 

$(2)$ KamLAND and JUNO are the LBL reactor-based antineutrino experiments having baselines more than 50 kilometers. These reactor experiments operate with $\bar{\nu}_e$ beam and look for the $\bar{\nu_e}$ appearance channel. In the limit $\theta_{13} \rightarrow 0$, the effective two flavour oscillation formula consists mainly the parameters $\Delta m_{21}^{2}$ and $\theta_{12} $.

$(3)$ Daya Bay is the short baseline (SBL) reactor experiment  having baseline of length of few kilometers only. Similar to LBL reactor experiments, Daya Bay also measures the disappearance channel, $\bar{\nu_e}\rightarrow\bar{\nu_e}$. Daya-Bay is almost unable to observe oscillations for small effective mass square difference, hence for $\Delta m_{21}^{2} \rightarrow 0$, it has sensitivity for $\Delta m_{31}^{2}$ and $\theta_{13}$ oscillation parameters.

\begin{table}[t]
	\centering
	\caption{Standard neutrino oscillation parameters and NSI parameters \cite{Coloma:2019mbs} in $1 \sigma$ interval. Here NSI parameters are assumed to be real.}
	\label{Tab1}
	\begin{tabular}{|c|c|}
		\hline
		Parameters  &  Best fit$\pm 1\sigma$\\
		\hline\hline
		$\theta_{12}^{o}$    &   $34.3\pm 1.0$\\
		\hline
		$\theta_{23}^{o}$    &   $48.79_{-1.25}^{+0.93}$\\
		\hline
		$\theta_{13}^{o}$    &   $8.58_{-0.15}^{+0.11}$\\
		\hline
		$\Delta m_{21}^{2}\times 10^{-5}\, \rm (eV^{2})$  &  $7.5_{-0.20}^{+0.22}$ \\
		\hline
		$\Delta m_{31}^{2}\times 10^{-3}\, \rm (eV^{2})$  &  $2.56_{-0.04}^{+0.03}$ \\
		\hline
		$\epsilon_{ee}$ & [0.24, 2.27] \\
		\hline
		$\epsilon_{\mu \mu}$ &[-0.30, 0.37] \\
		\hline
		$\epsilon_{\tau \tau}$ & [-0.30, 0.38] \\
		\hline
		$\epsilon_{e \mu}$ & [-0.33, 0.16] \\
		\hline
		$\epsilon_{e \tau}$ & [-0.76, 0.53] \\
		\hline
		$\epsilon_{\mu \tau}$ &[-0.03, 
		0.03] \\
		\hline
	\end{tabular}
	
\end{table}

DUNE is the LBL accelerator experiment which identifies mainly $\nu_{\mu}\rightarrow \nu_{e}$ channel. However, since DUNE also has high potential to detect $\nu_{\tau}$ events \cite{DeGouvea:2019kea, Machado:2020yxl}, here we have considered $\nu_{\mu}\rightarrow \nu_{\tau}$ flavour transition in case of DUNE. For MINOS and T2K also,  we have  considered  $\nu_{\mu}\rightarrow \nu_{\tau}$ oscillation channel \cite{MINOS:2006foh}.

We obtain  results for two flavour neutrino oscillations in three different scenarios, {\it i.e.,} when neutrino oscillations occur: (i) in vacuum, (ii) under the effects of SM interaction and (iii) with NSI effects. The values of the NSI parameters are taken from the ref. \cite{Coloma:2019mbs} in which the parameters are constrained by accounting the $CP$-conserving variables that results in real NSI parameters. The values of standard neutrino oscillation parameters \cite{deSalas:2020pgw} along with NSI parameters \cite{Coloma:2019mbs} are given in Table \ref{Tab1}. The analytical expressions of the two parameters, $M(\rho)$ and $N_{l_1}(\rho)$, are obtained as
\begin{figure*}[ht!]
\includegraphics[scale=0.38]{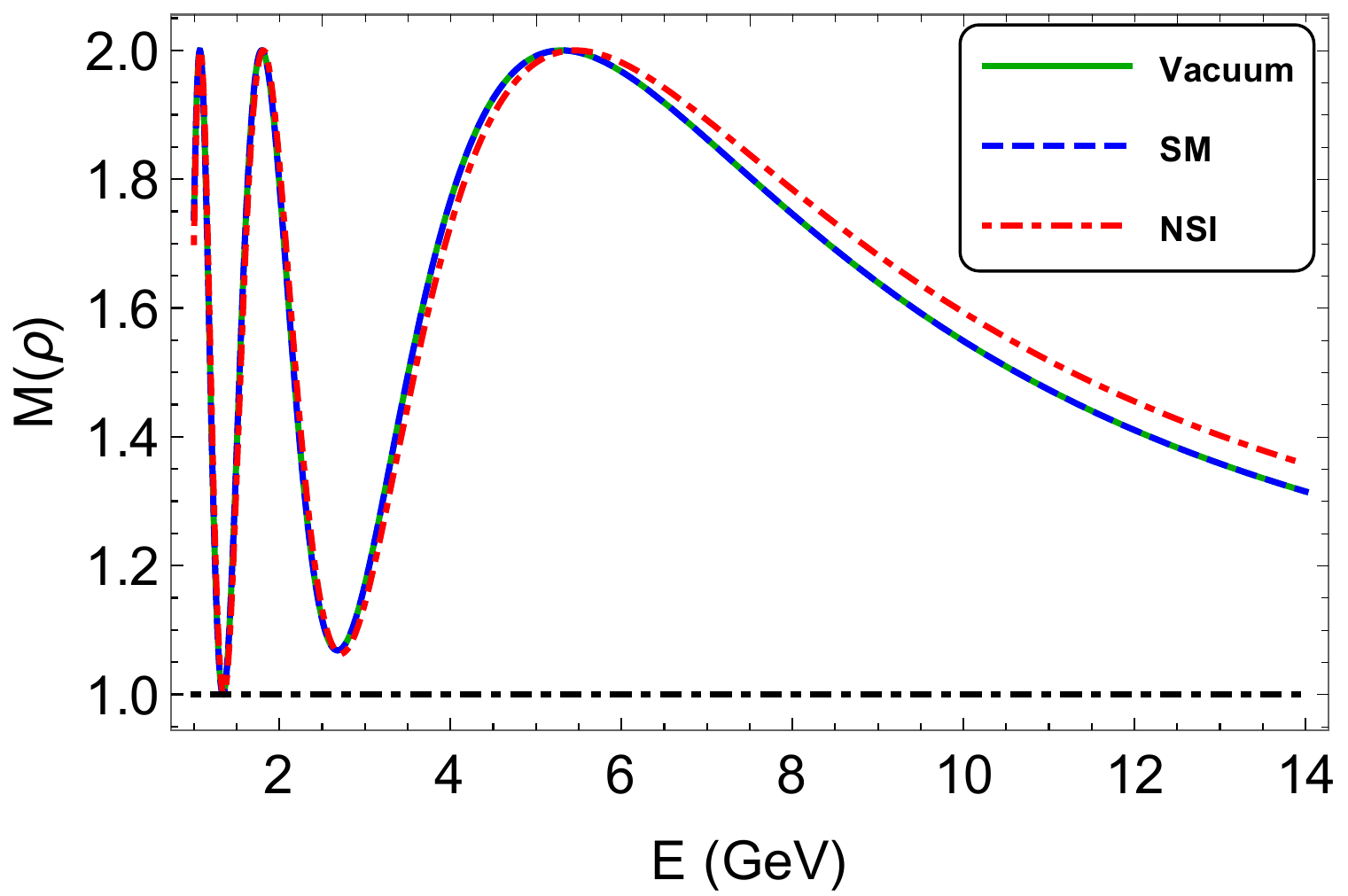}
\includegraphics[scale=0.38]{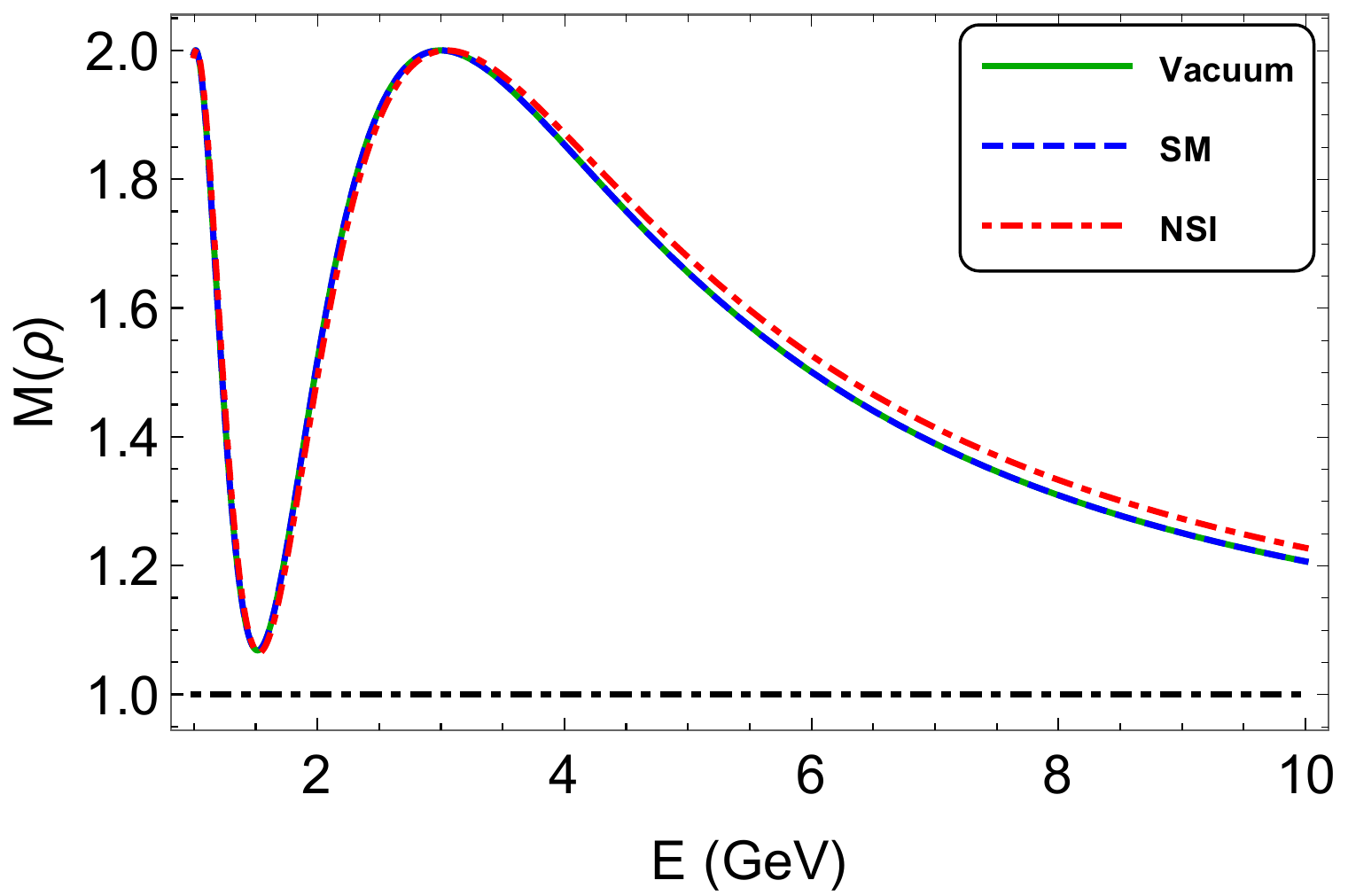}
\includegraphics[scale=0.38]{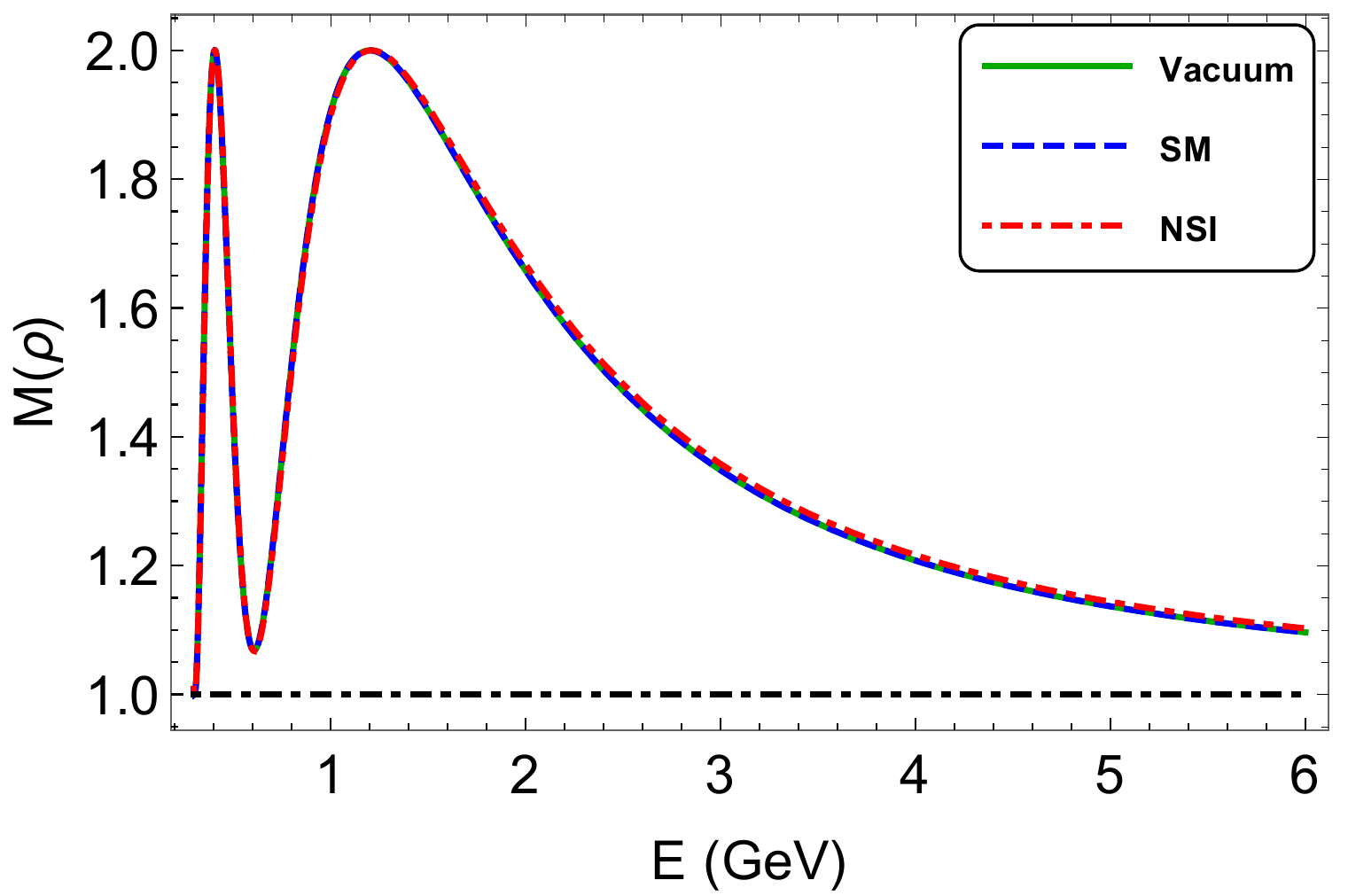}\\
\includegraphics[scale=0.38]{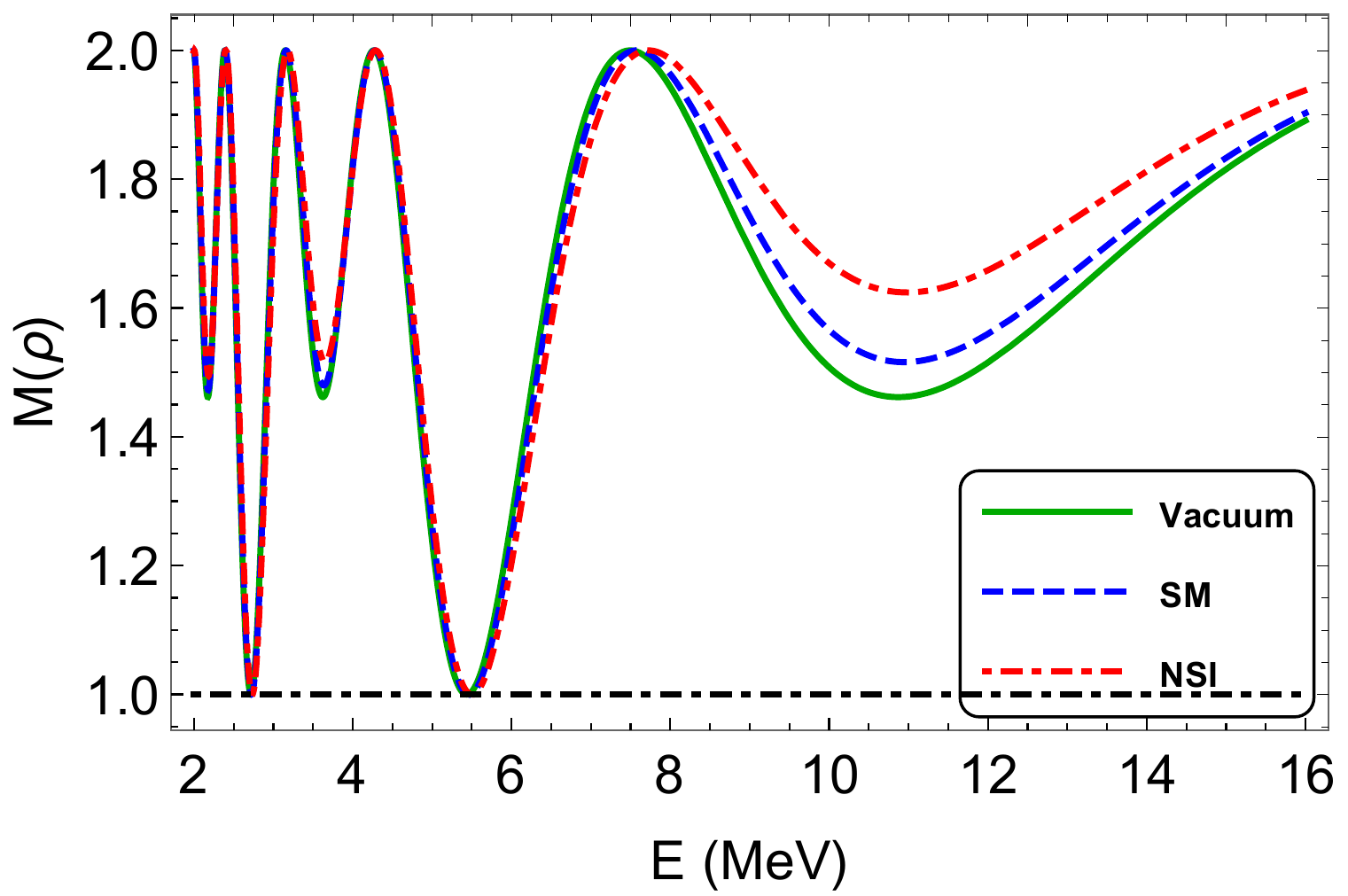}
\includegraphics[scale=0.38]{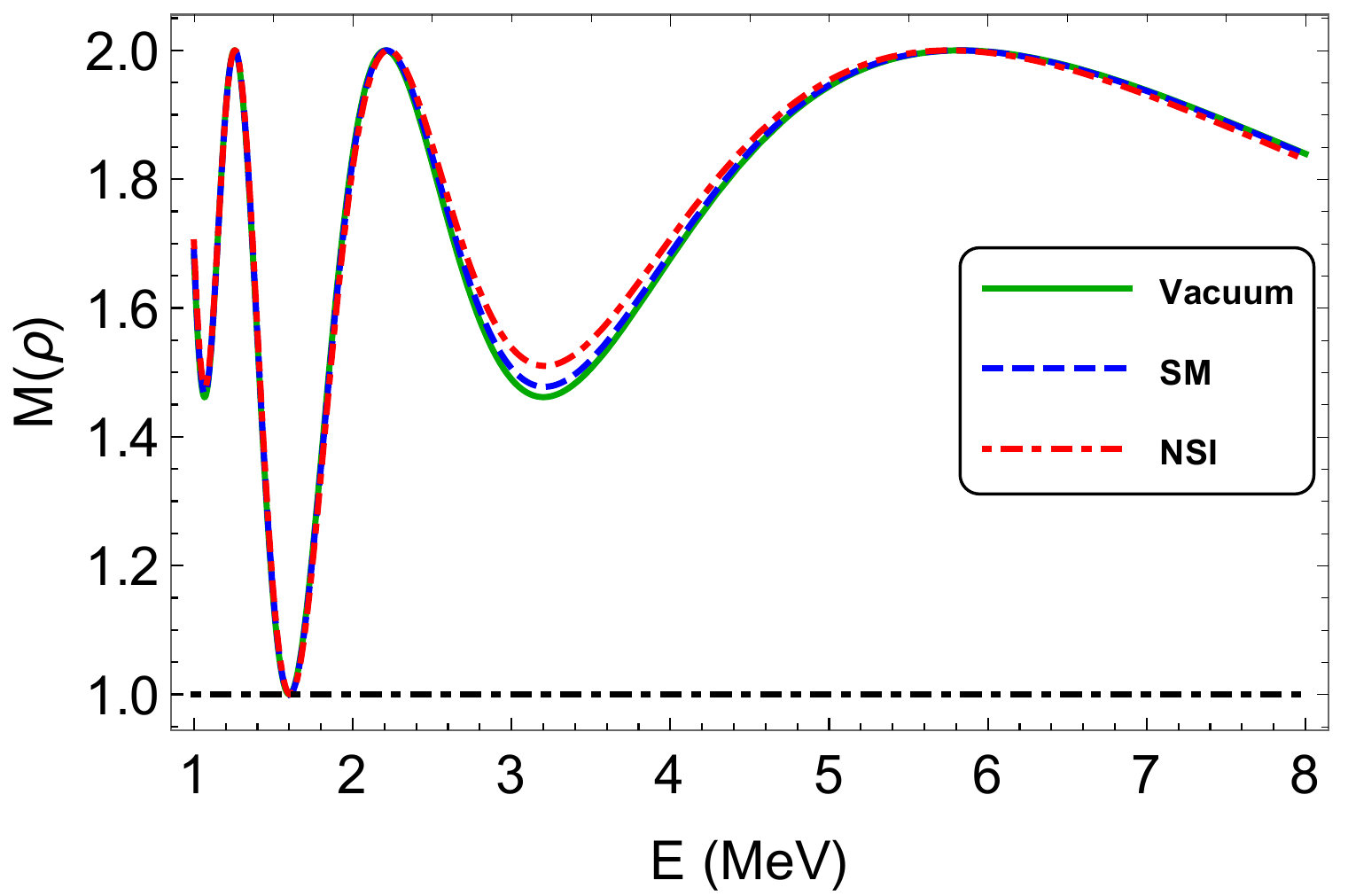}
\includegraphics[scale=0.38]{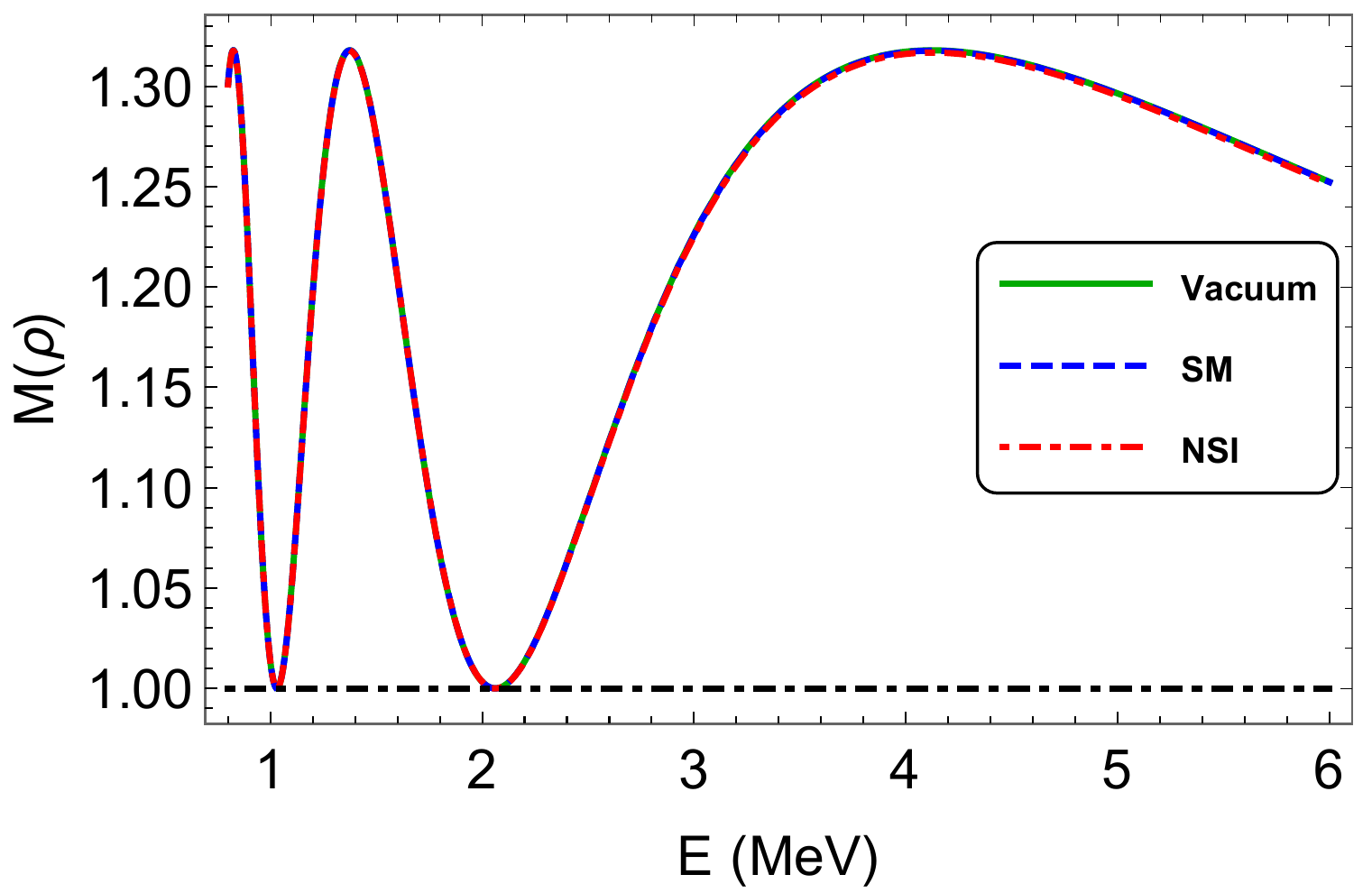}
\caption{\ Variation of M($\rho$) with energy ($E$) for the accelerator and reactor experiments: (a) Upper left: DUNE, $L=1300$ km, $E\approx1-14$ GeV; (b) upper middle: MINOS, $L=735$ km, $E\approx1-10$ GeV; (c) upper right: T2K, $L=295$ km, $E\approx0-6$ GeV;  (d) lower left: KamLAND, $L=180$ km, $E\approx1-16$ MeV; and (e) lower middle: JUNO, $L=53$ km, $E\approx1-8$ MeV; (f) lower right: Daya Bay, $L=2$ km,  $E\approx0.8-6$ MeV. The solid (green) line corresponds to oscillation in vacuum, the dashed (blue) and dot-dashed (red) lines represent the results for SM and NSI, respectively. Dotted (black) line represents the classical bound of M($\rho$).}
\label{fig1}
\end{figure*}


\begin{eqnarray}
  M\left ( \rho  \right )=f_a(x,y,r)+f_b(x,y,z,r)+f_c(x,z,r), \label{mrho}\\
  N_{l_1}(\rho) =  2+\sqrt{\frac{2 f_b(x,y,z,r)}{3}},~~~~~~~~~~~~~~~~~~~~~~\label{Nl1}
\end{eqnarray}
where the form of quantities $f_a$, $f_b$ and $f_c$ are given as
\small
\begin{eqnarray}
 f_a(x,y,r)&=&\frac{e^{{\rm Im}(4r)}[x^{2}+y^{2}+(x^{2}-y^{2})\cos(2r)]^{2}}{4x^{4}},\nonumber\\
 f_b(x,y,z,r)&=&\frac{3e^{{\rm Im}(4r)}z^{2}\sin^{2}r(x^{2}+y^{2}+(x^{2}-y^{2})\cos(2r))}{x^{4}},\nonumber\\
 f_c(x,z,r)&=&\frac{e^{{\rm Im}(4r)}z^{4}\sin^{4}r}{x^{4}},
\label{r2}
\end{eqnarray}
\normalsize
with $r=\frac{L x}{4E}$. The quantities $x$, $y$ and $z$ are functions of NSI parameters and are given by
\begin{eqnarray}\label{eqxyz}
x =\sqrt{x_{1}-x_{2}+x_{3}},~~~~~~~~~~~~~~~~~~~~~\notag\\
y = \frac{-x_2}{2\,\Delta m^{2}\,\cos(2\theta)}+\Delta m^{2}\,\cos(2\theta),\\
z = \frac{x_3-(\Delta m^{2})^{2} \cos(4\theta)}{2\,\Delta m^{2}\,\sin(2\theta)},~~~~~~~~~~~~~\notag  
\end{eqnarray}
with 
\begin{eqnarray}
x_{1}= 4A^{2}E^{2}(4\epsilon _{\alpha \beta }^{2}+(\epsilon _{\alpha \alpha }+b-\epsilon _{\beta \beta })^{2}),~~\notag\\
x_{2}=4\,A\,E\,(\epsilon _{\alpha \alpha }+b-\epsilon _{\beta \beta })\,\Delta m^{2}\cos(2\theta ),\\
x_{3}=8\,A\,E\,\epsilon _{\alpha \beta }\,\Delta m^{2}\,\sin(2\theta )+(\Delta m^{2})^{2}.~~\notag
\label{rr4}
\end{eqnarray}

In Eq. \eqref{eqxyz}, the quantities $x$, $y$, $z$ vary for different experimental set-ups due to the presence of distinct NSI parameters corresponding to separate oscillation channels. In Eq. \eqref{rr4}, the parameter $b$ represents the SM matter effect {\it i.e.,} we have $b = 1$ when $\bar{\nu}_e $ participates in transitions and $b = 0$ for ${\nu}_\mu\rightarrow {\nu}_{\tau}$ transition where $\bar{\nu}_e $ is not included. The parameter $x$ given in Eq. \eqref{eqxyz} represents the effective mass square difference in the presence of NSI for two flavour neutrino oscillations. Similarly, in presence of NSI, the effective mixing angle can be given by
\begin{eqnarray}\label{eff}
\sin 2\theta_{NSI} = \frac{z}{x},\\
\cos 2\theta_{NSI} = \frac{y}{x}.
\end{eqnarray}
The solutions for the oscillation in vacuum are restored if the matter potential $A$ and NSI parameters $\epsilon_{\alpha\beta}$ are set to zero.

NAQC parameter, $N_{l_1}(\rho)$, represents the measure of quantumness in terms of entanglement and steerability in a given composite system. This can be seen by the definition of NAQC given as $2 + \mathcal{C}$ \cite{Ming:2020nyc}, where $\mathcal{C}$ depicts the concurrence, a well defined measure of entanglement for two-level systems. Concurrence can be represented in terms of of transition probabilities as follows \cite{ALOK201665}
\begin{equation}
	\mathcal{C} = 2\sqrt{P_{sur}P_{osc}},
\end{equation}
{\it i.e.,} both $\mathcal{C}$ and $N_{l_1}(\rho)$ can be treated as functionals of  survival ($P_{sur}$) and oscillation probabilities ($P_{osc}$). 

In presence of NSI, the algebraic expressions for $P_{sur}$ and $P_{osc}$ in terms of the parameters $x$, $y$, $z$ and $r$ are expressed as
\begin{eqnarray}
  P_{sur}&=& \frac{e^{{\rm Im}(2r)}\left[x^{2}+y^{2}+(x^{2}-y^{2})\cos(2r)\right]}{2x^{2}}, \nonumber\\
 P_{osc}&=&\frac{e^{{\rm Im}(2r)}z^{2}\sin^{2}(r)}{x^{2}}.
\end{eqnarray}
The factors $f_a$, $f_b$ and $f_c$ are represented in terms of $P_{sur}$ and $P_{osc}$ below for convenience
\begin{equation}
f_a=P_{sur}^2, \hspace{1mm} f_b=6P_{surv}P_{osc}, \hspace{1mm} f_c=P_{osc}^2\,.
\end{equation}

\begin{figure*}
	\includegraphics[scale=0.38]{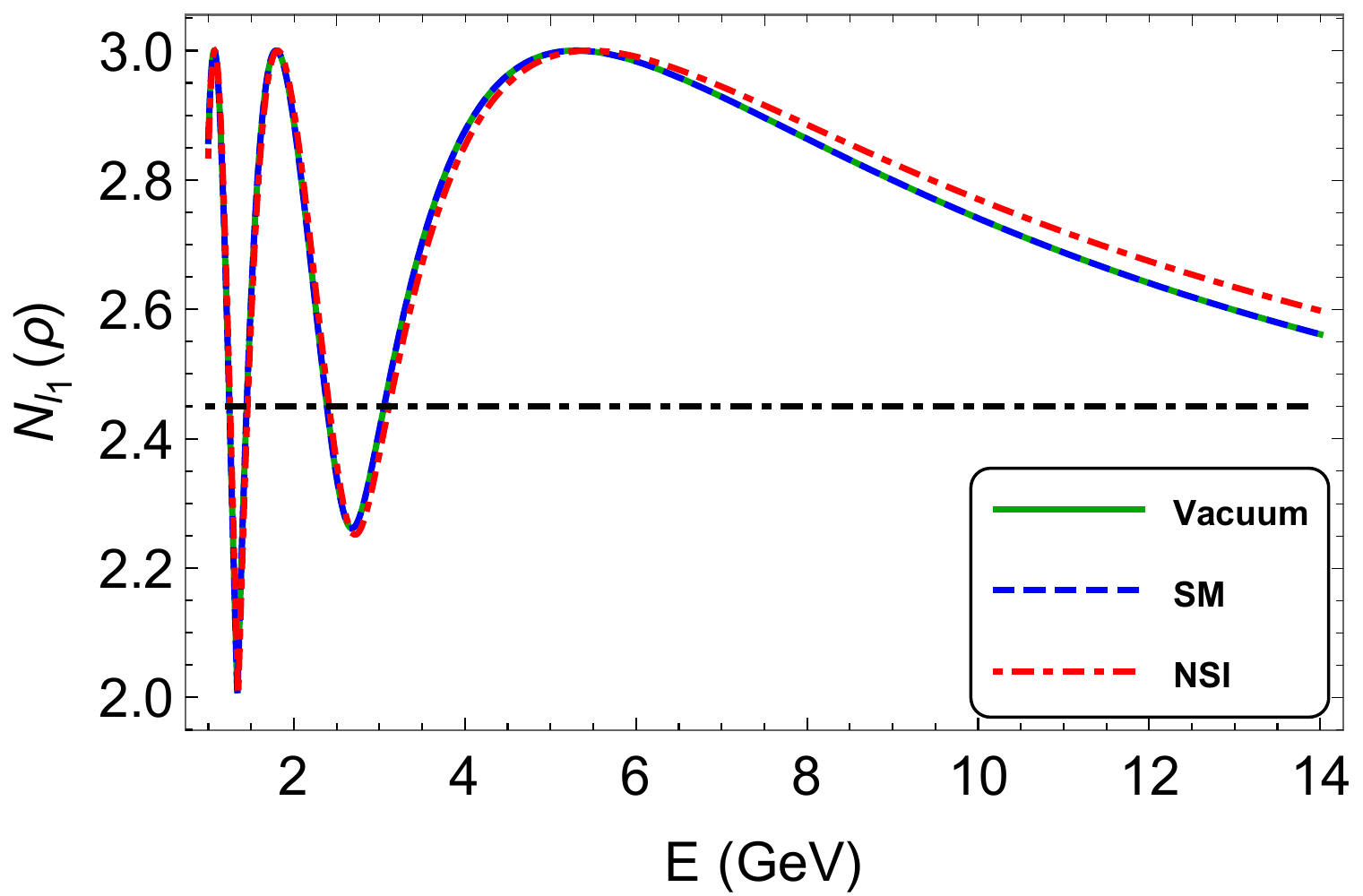}
	\includegraphics[scale=0.38]{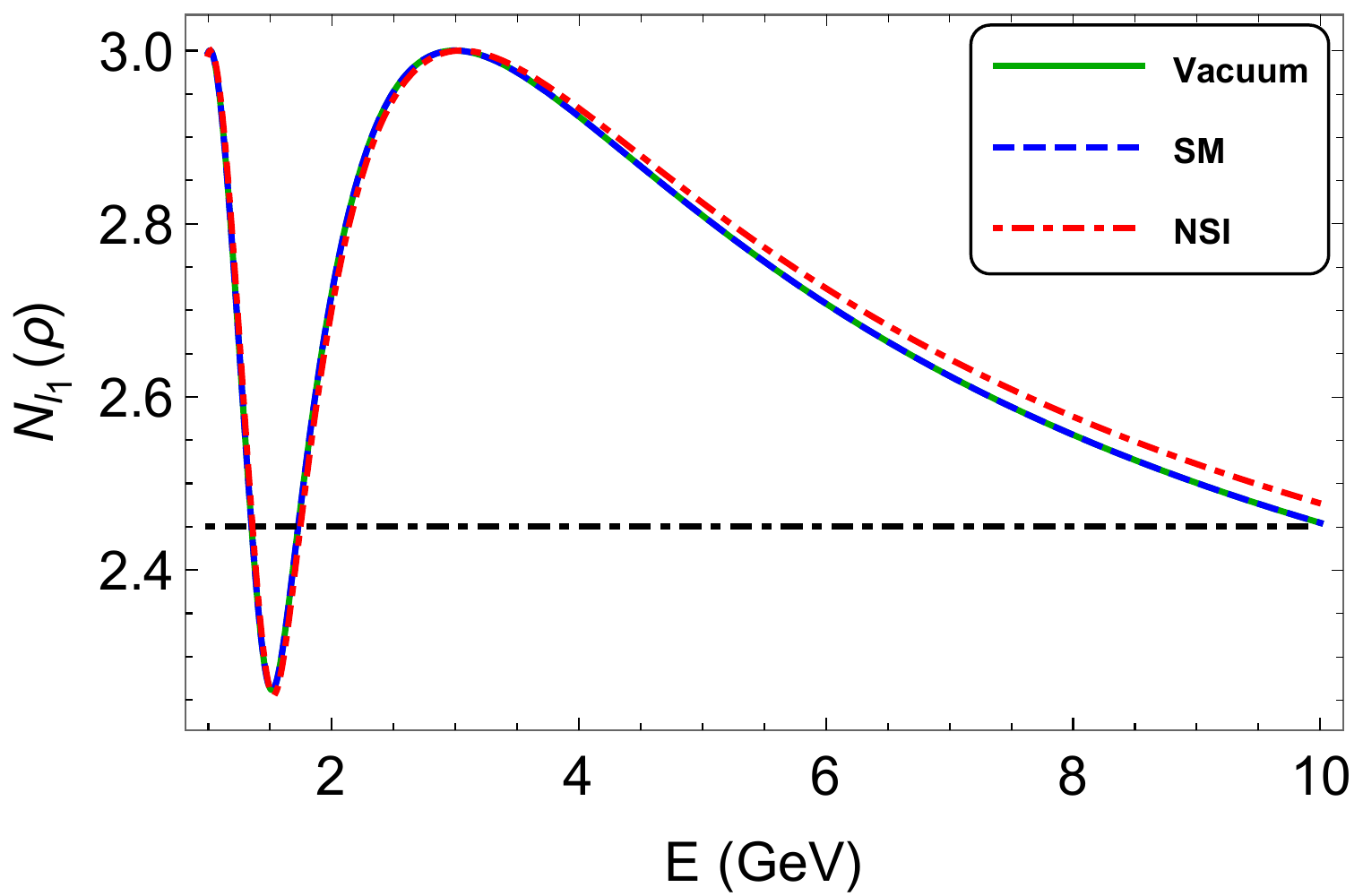}
	\includegraphics[scale=0.38]{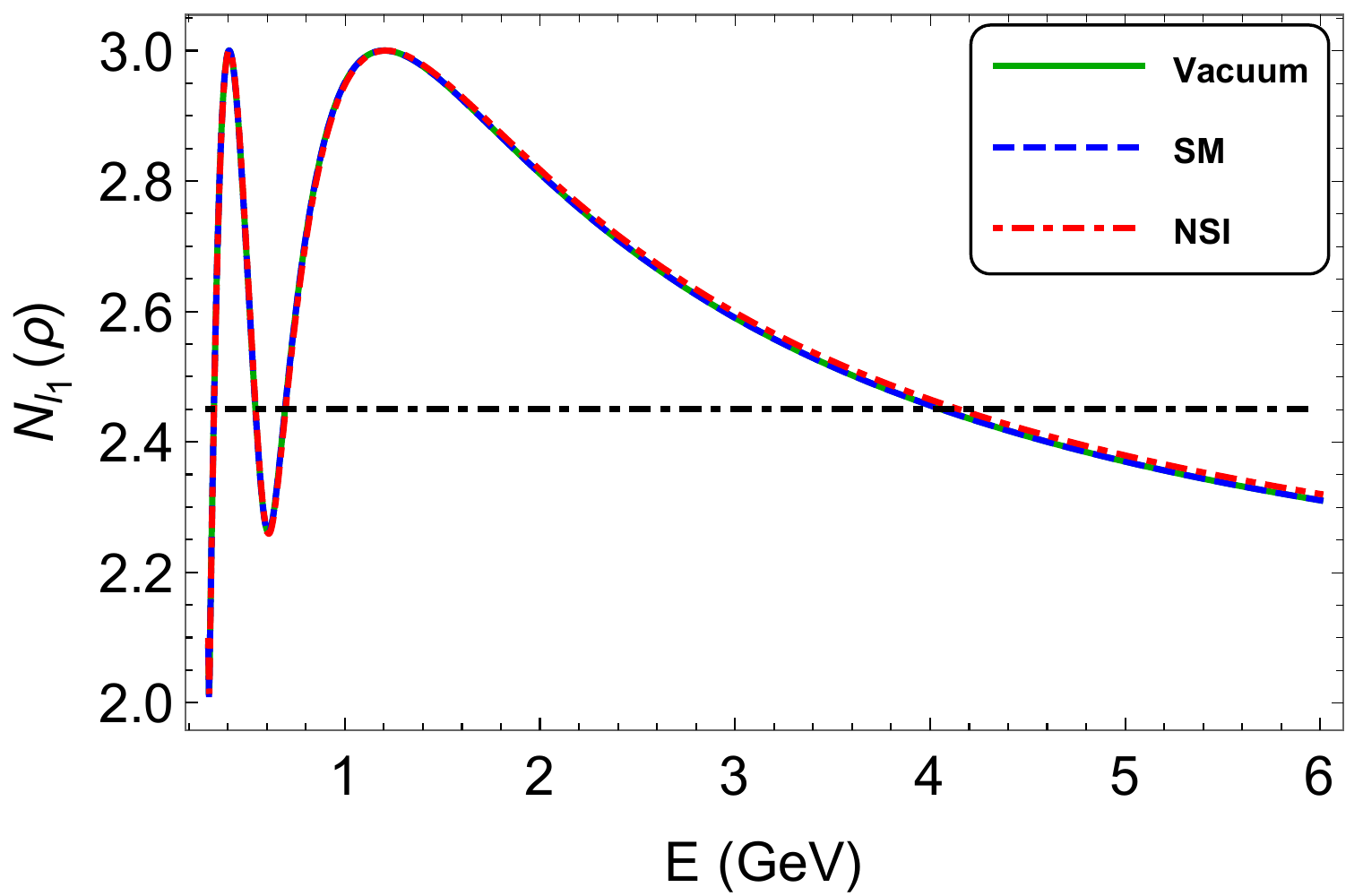}\\
	\includegraphics[scale=0.38]{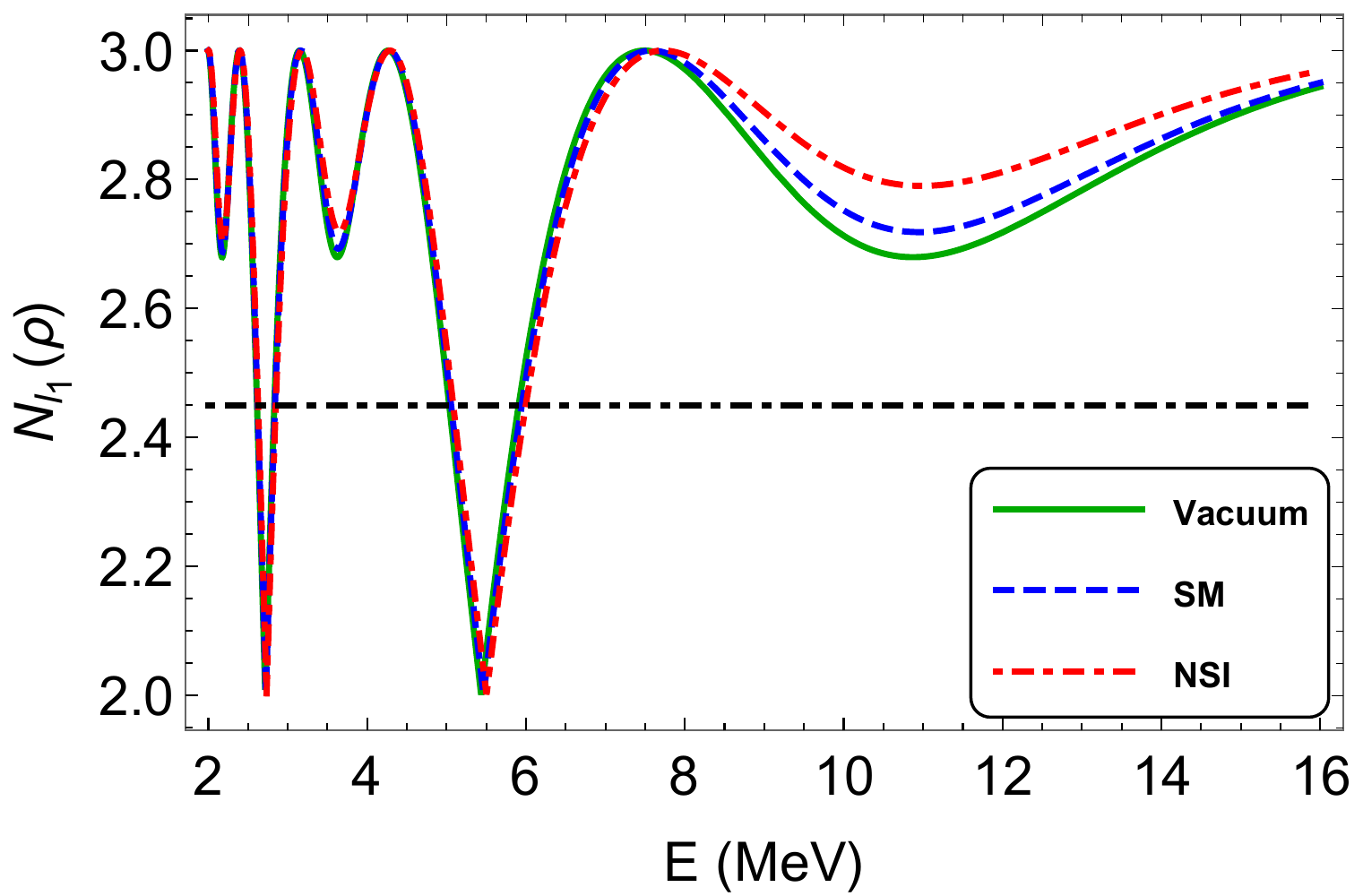}
	\includegraphics[scale=0.38]{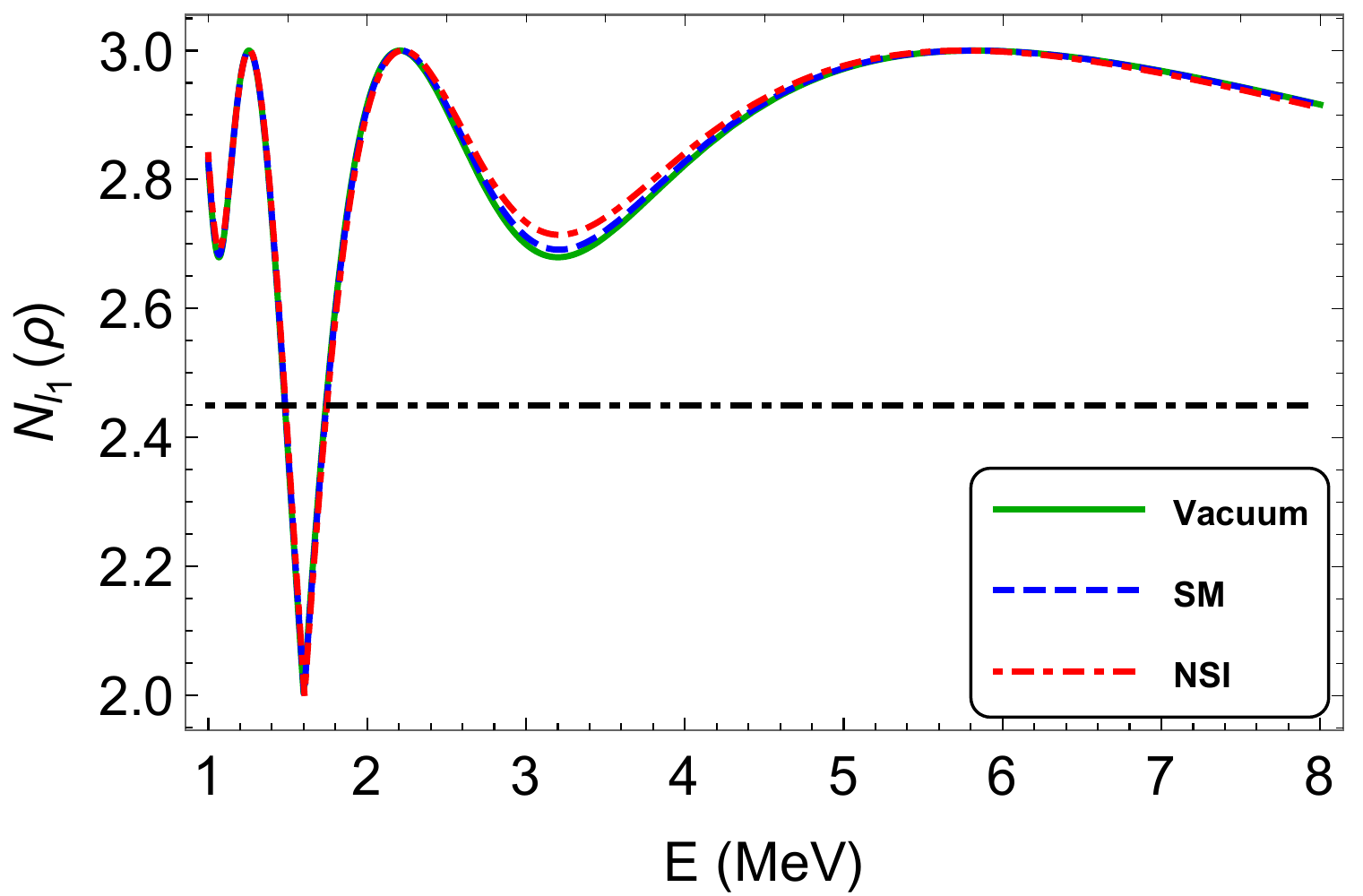}
	\includegraphics[scale=0.38]{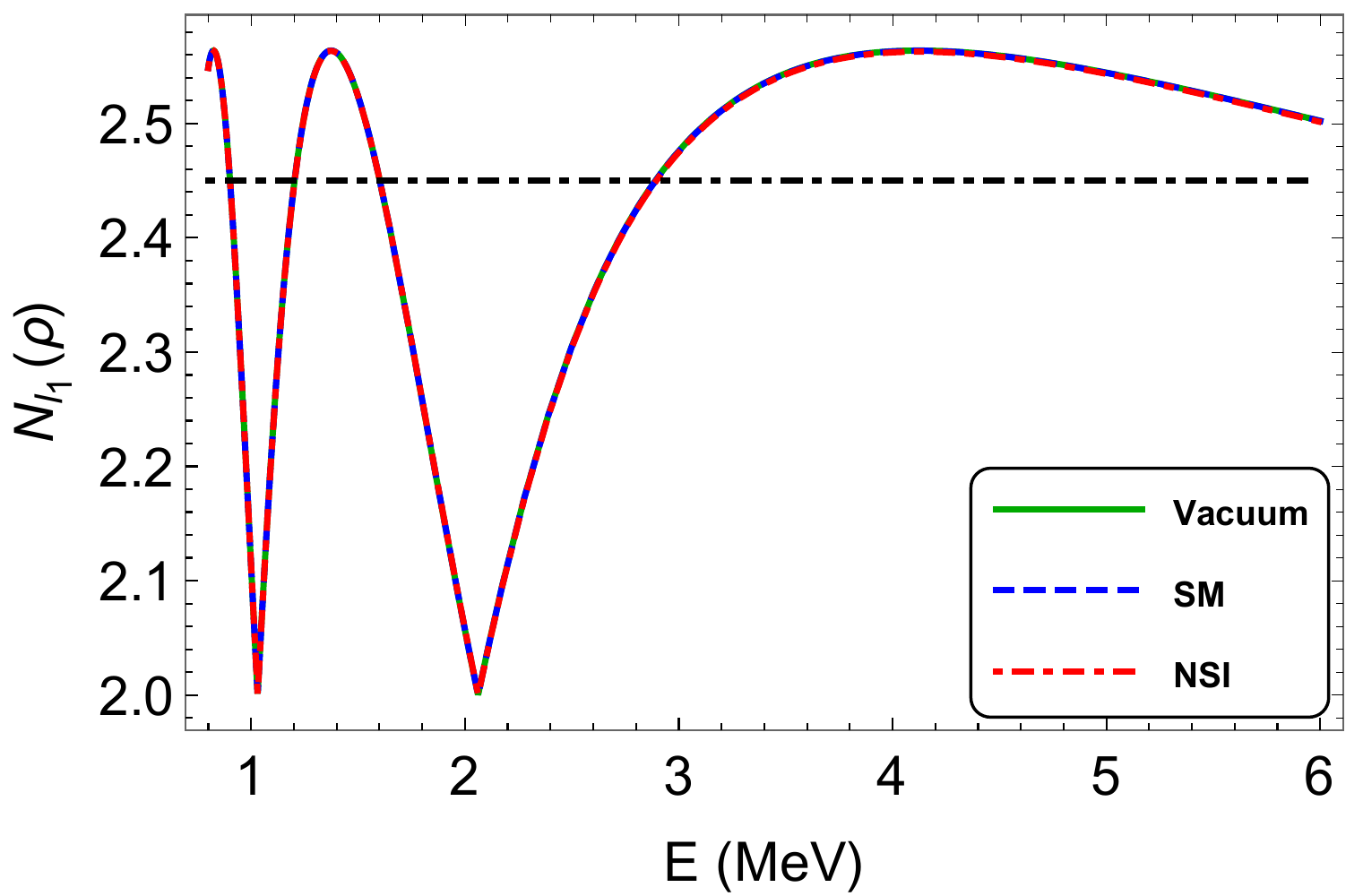}
	\caption{Variation of NAQC parameter
		with energy ($E$) for the accelerator and reactor experiments: (a) Upper left: DUNE, $L=1300$ km, $E\approx1-14$ GeV; (b) upper middle: MINOS, $L=735$ km, $E\approx1-10$ GeV; (c) upper right: T2K, $L=295$ km, $E\approx0-6$ GeV;  (d) lower left: KamLAND, $L=180$ km, $E\approx1-16$ MeV; and (e) lower middle: JUNO, $L=53$ km, $E\approx1-8$ MeV; (f) lower right: Daya Bay, $L=2$ km, $E\approx0.8-6$ MeV. The solid (green) curve corresponds to oscillation in vacuum, the dashed (blue) and dot-dashed (red) curves represent the results for SM and NSI,  respectively. Dotted (black) line represents the upper bound of NAQC \cite{PhysRevA.95.010301}.}
	\label{fig2}
\end{figure*}

In Fig. (\ref{fig1}) and (\ref{fig2}), we present the comparative analysis of $M(\rho)$ and $N_{l_1}(\rho)$, respectively for two flavour neutrino oscillations in  vacuum, SM  and with NSI. In all of these plots, the solid line (green) corresponds to the results for oscillations in vacuum, while the dashed (blue) and dot-dashed (red) lines represent SM and NSI predictions,  respectively. The result for $M(\rho)$ in the case of three accelerator experiments are illustrated in the  upper panels of Fig. (\ref{fig1}) with DUNE in upper left panel, MINOS in the upper middle panel and T2K in the upper right panel. The same are displayed in Fig. (\ref{fig2}) for $N_{l_1}(\rho)$. Bell's inequality violation is realized when the value of $M(\rho)$ exceeds 1, while NAQC is violated when the value of $N_{l_1}(\rho)$ becomes greater than $\approx$ 2.45. 

It can be seen that the Bell's inequality violation occurs in the entire energy range for all three accelerator experiments, while the NAQC violation  observed only in some specific ranges of energy. This justifies the more elementary feature of $N_{l_1}(\rho)$ compared to $M(\rho)$ which implies that Bell's inequality has to be violated if NAQC violation occurs.  For example, in case of DUNE, NAQC is violated within a narrow energy range around $E = 1$ and $2$ GeV and also in a wider energy range  $3 \leq E \leq 14$ GeV, while Bell's inequality violation is noticed in the entire energy range ($1 \leq E \leq 14$ GeV). The energy-intervals showing violations of these two measures are given in Table \ref{Tab2} for all the six experiments considered in this work. These intervals are calculated for the case of NSI. However, these intervals remain the same even for oscillations in vacuum and in matter with SM interaction. Thus we find that the NAQC would always be violated in these energy ranges irrespective of the type of interaction.

\begin{table}[h]
	\centering
	\caption{Energy regions showing the violation of $M\rho$ and NAQC for six different experimental set-ups. For accelerator experiments the energy range lies in GeV region, while for reactor experiments they are in MeV.  }
	\label{Tab2}
	\begin{tabular}{|c|c|c|}
		\hline
		Experiments& $ N_{l_1}(\rho)$  & $M(\rho) $ \\
		\hline\hline
		DUNE (GeV)  & 1-1.25  ,1.5-2.45  ,3-14& 1-14  \\
		\hline
		MINOS (GeV)   & 1-1.3 ,1.75-10   & 1-10 \\
		\hline
		T2K (GeV)   &0.4-0.5 , 0.7-4 &  0-6 \\
		\hline
		KamLAND ( MeV)  & 1-2.5 , 3-5 ,6-16 & 1-16 \\
		\hline
		JUNO (MeV)   &1-1.45 , 1.8-8  & 1-8 \\
		\hline
		Daya Bay (MeV)   & 0.8-0.9  ,1.2-1.6  ,2.9-6 &0.8-6 \\
		\hline
		
	\end{tabular}
	
\end{table}

Analysing the nonclassical features represented in terms of $M(\rho)$ and $N_{l_1}(\rho)$ in the upper panels of Fig. (\ref{fig1}) and (\ref{fig2}) respectively, for the accelerator experiments, it can be seen that the effect of NSI is more conspicuous in case of the experiments with longer baseline and higher neutrino-energy range. Therefore maximum effect of NSI is visible in case of DUNE with $L=1300$ km. MINOS ($L=735$ km) shows smaller deviation in case of NSI compared to DUNE, while T2K shows no significant change in the correlation quantities, as it has the shortest baseline ($L=295$ km) and lower neutrino-energy. Both DUNE and MINOS exhibit the NSI effect in the higher side of energy range ($\gtrsim 4$ GeV). 

\begin{figure*}
	\centering
	\includegraphics[scale=0.45]{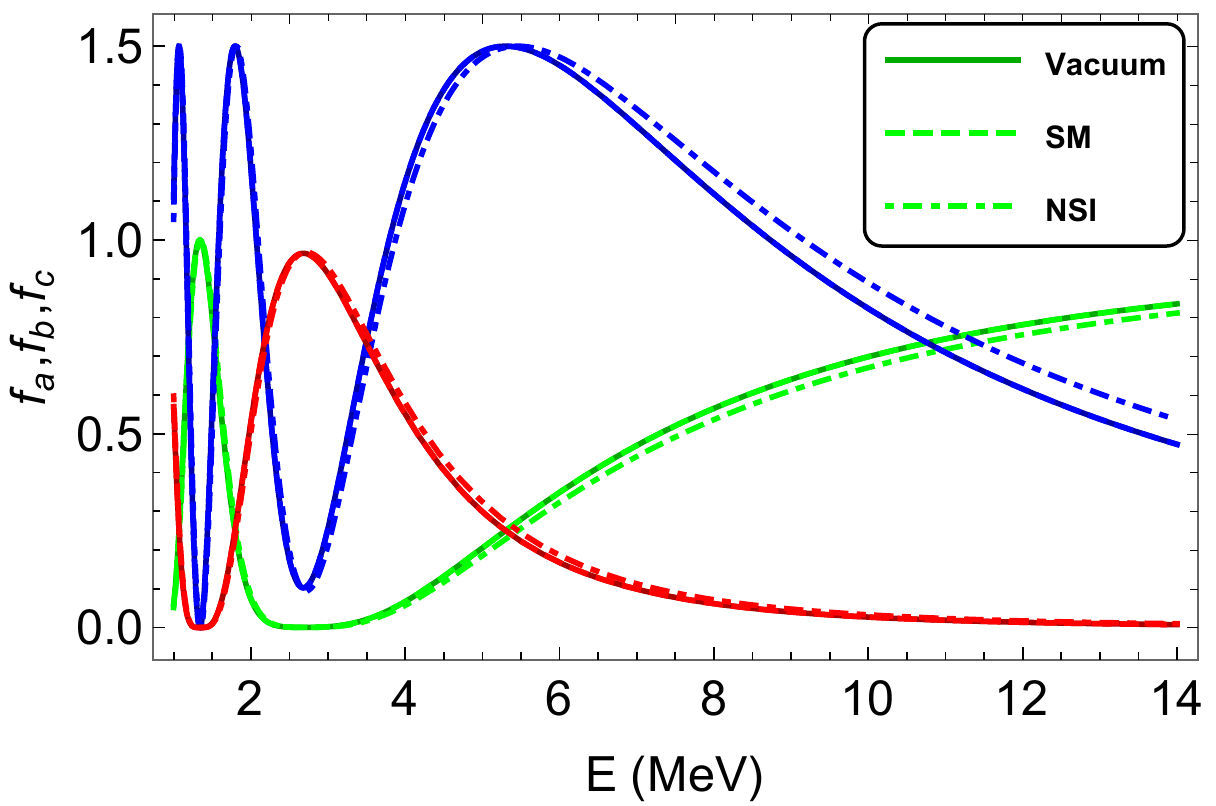}
	\includegraphics[scale=0.45]{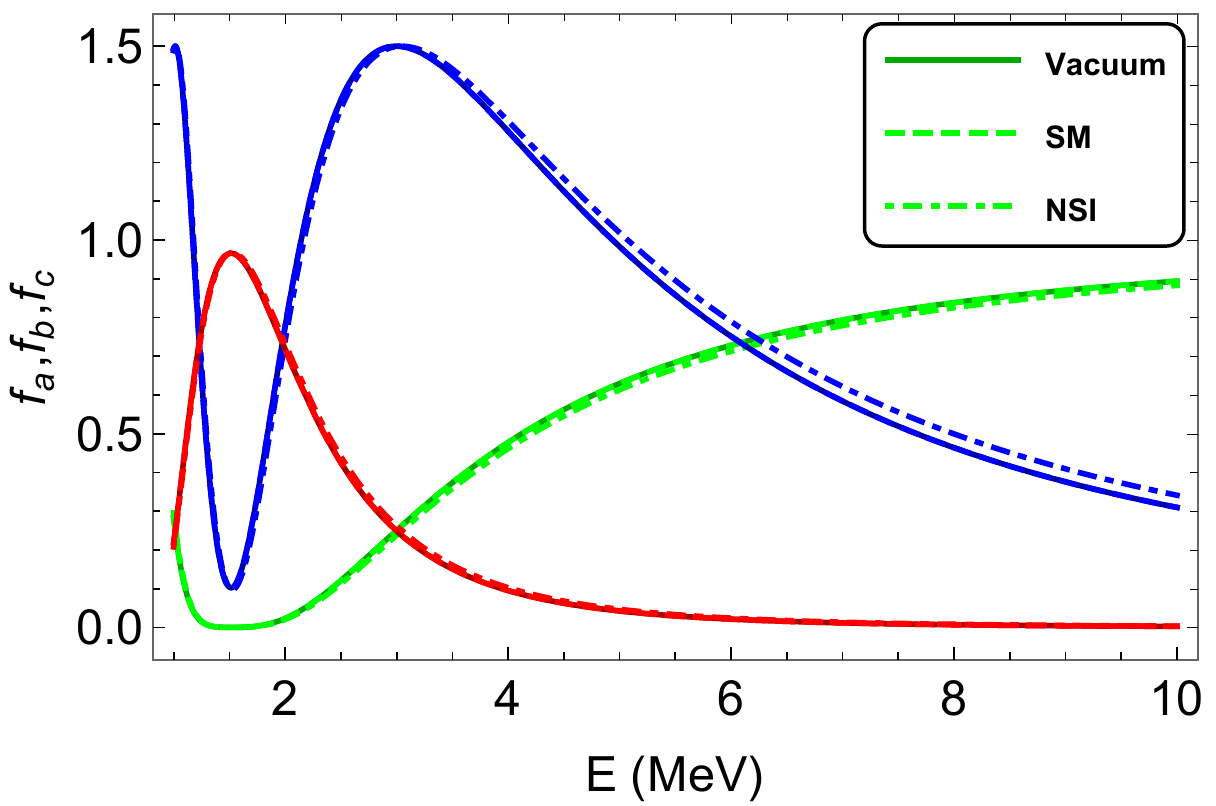}
	\includegraphics[scale=0.45]{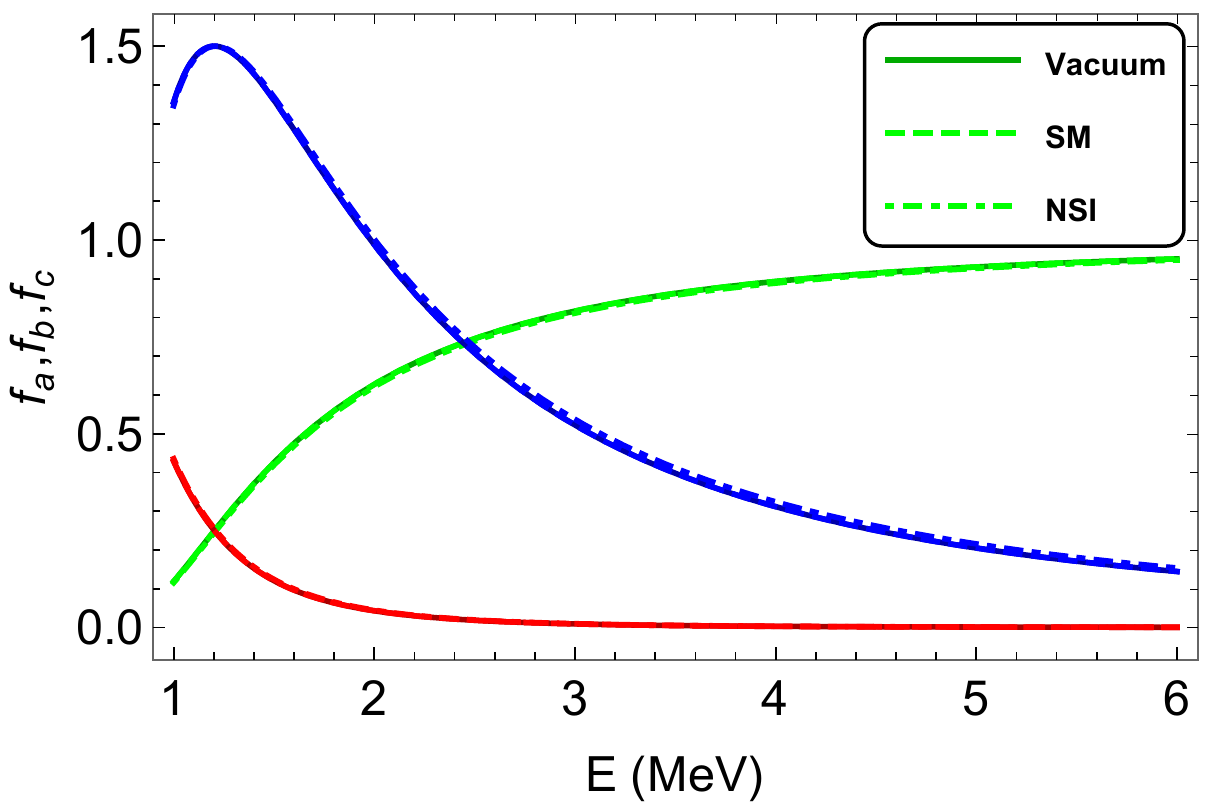}\\
	\includegraphics[scale=0.45]{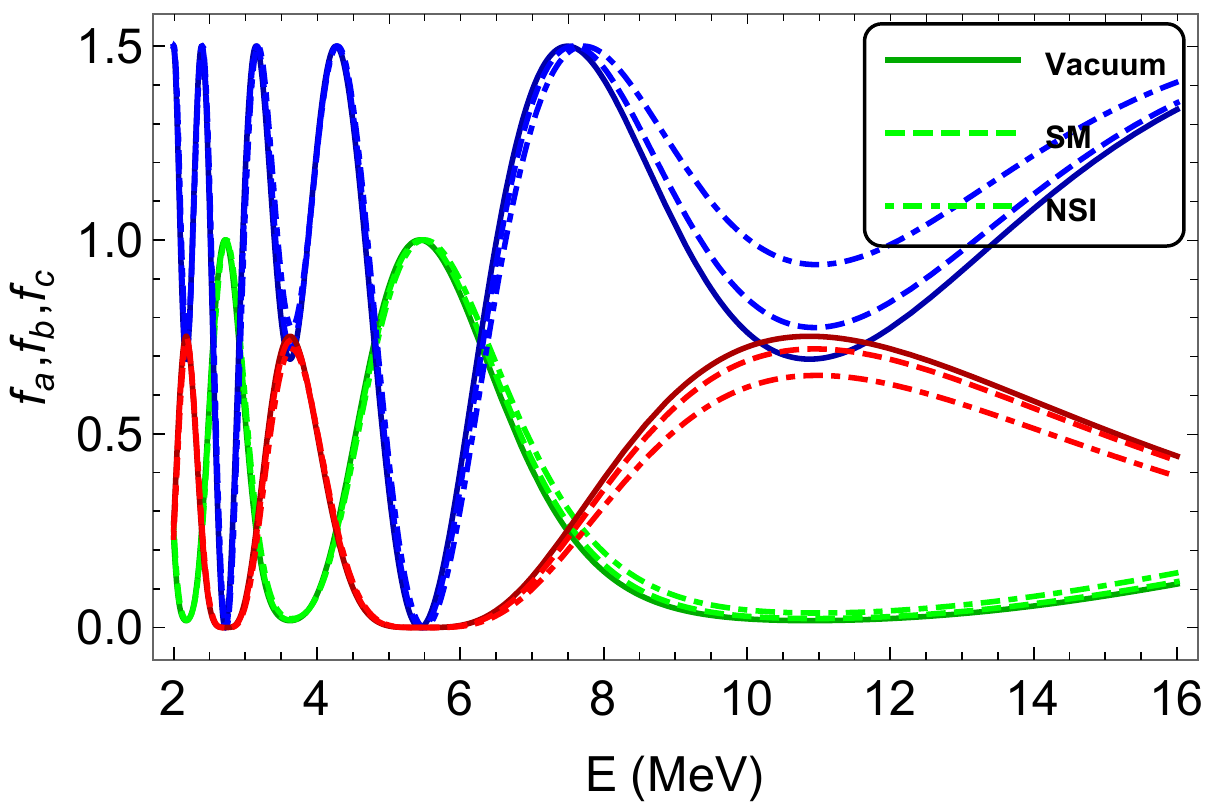}
	\includegraphics[scale=0.45]{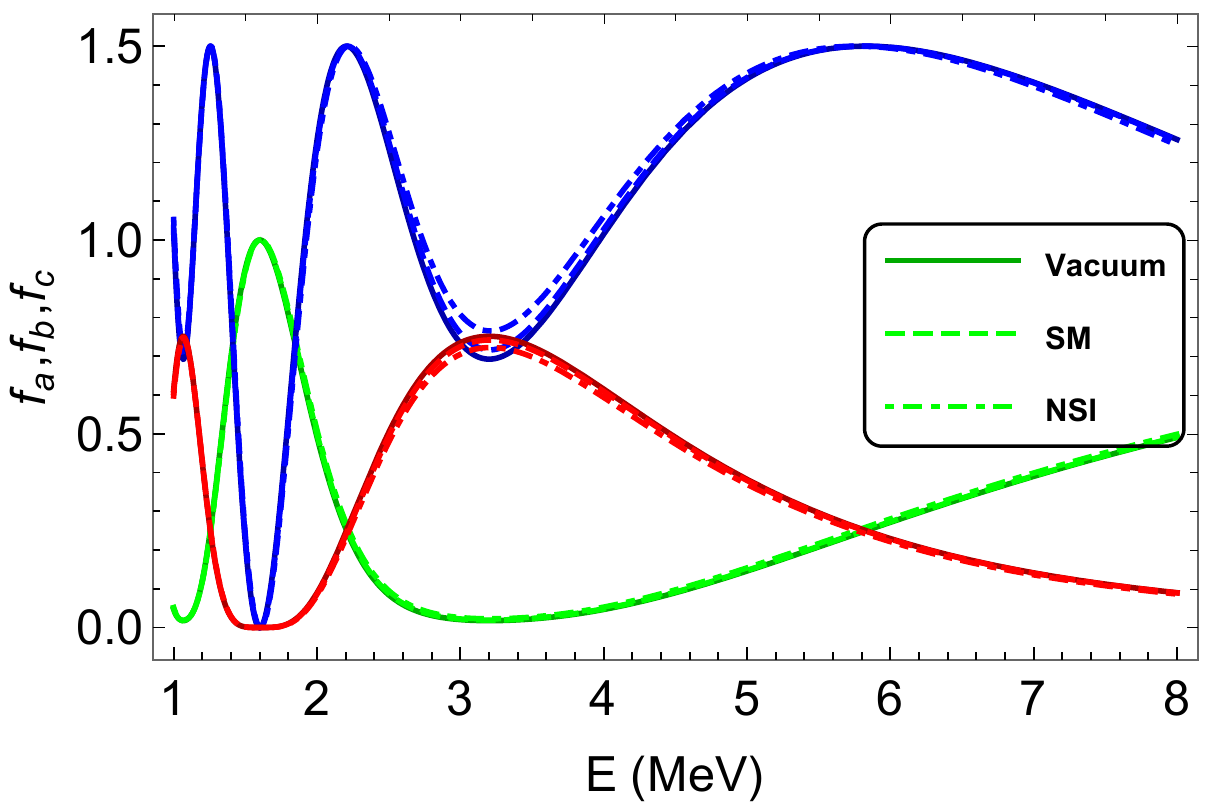}
	\includegraphics[scale=0.45]{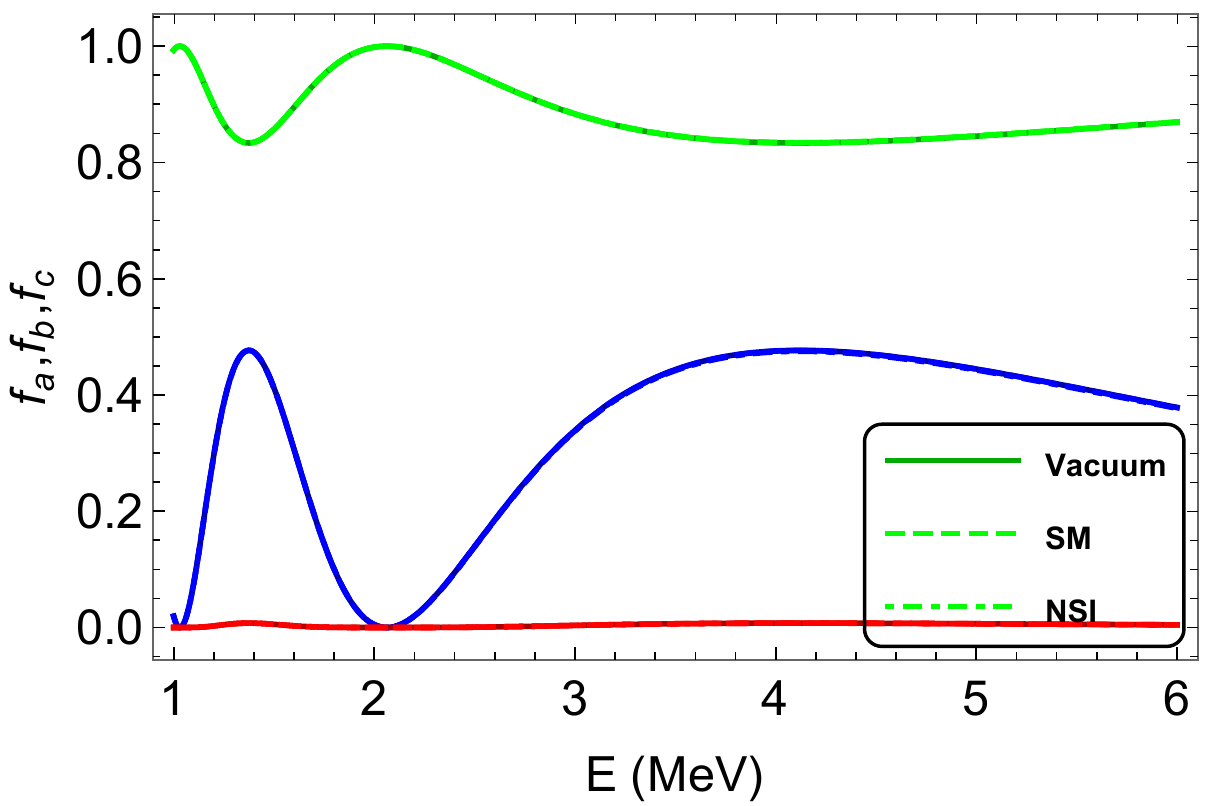}
	
	\caption{Variation of  $f_{a}$ (green), $f_{b}$ (blue) and $f_{c}$ (red) with energy ($E$) for the accelerator and reactor experiments:  (a) Upper left: DUNE, $L=1300$ km, $E\approx1-14$ GeV; (b) upper middle: MINOS, $L=735$ km, $E\approx1-10$ GeV; (c) upper right: T2K, $L=295$ km, $E\approx0-6$ GeV;  (d) lower left: KamLAND, $L=180$ km, $E\approx1-16$ MeV; and (e) lower middle: JUNO, $L=53$ km, $E\approx1-8$ MeV; (f) lower right: Daya Bay, $L=2$ km, $E\approx0.8-6$ MeV. The solid, dashed and dot-dashed curves correspond to neutrino oscillations under vacuum, SM and NSI scenarios, respectively.}
	\label{fig3}
\end{figure*}


In the lower panel of Fig. (\ref{fig1}) and (\ref{fig2}) we present the plots of $M(\rho)$ and $N_{l_1}(\rho)$ respectively, for the reactor experiments. In both the figures the lower left panel corresponds to KamLAND experiment ($L=180$ km), the lower middle panel is for JUNO ($L=53$ km) and the lower right panel shows the result for Daya Bay ($L=2$ km). Similar to accelerator experiments, we find the violation in $M(\rho)$ over the entire energy range of the reactor experiments, while $N_{l_1}(\rho)$ is violated in certain energy ranges only, given in Table \ref{Tab2}. The effect of NSI is maximally observed in the case of KamLAND experiment, which has the longest baseline, in the energy region $8\leq E \leq 15$ MeV that peaks at $E\sim 10.5$ MeV. The effect of NSI becomes much smaller in case of JUNO at $E\sim 3$ MeV, while it becomes completely negligible for the Daya Bay experiment. KamLAND gives the opportunity to probe NSI at higher energies ($\geq 8$ MeV). Although the reactor neutrinos in KamLAND can be upto $E\sim 10$ MeV, solar neutrinos are detectable within the range $10$ MeV $\leq E \leq 15$ MeV  \cite{Perevozchikov:2009bla}. In our analysis we have also considered solar neutrino parameters in the case of KamLAND experiment.

\begin{table}
	\caption{Percentage (\%) increase in Bell's inequality parameter $M(\rho)$ and NAQC in presence of NSI in comparison to vacuum and SM interaction for six different experimental set-ups.}
	\begin{tabular}{|c|c|c|c|}
		\hline
		Expts. & Measure & \% inc. w.r.t. vac & \% inc. w.r.t. SM int. \\
		\hline\hline
		DUNE & $M(\rho)$ & 4.3 & 4.3 \\
		\hline
		& $N_{l_1}(\rho)$ & 4 & 4  \\
		\hline 
		MINOS & $M(\rho)$ & 2.4 & 2.4  \\
		\hline
		& $N_{l_1}(\rho)$ & 2.3 & 2.3  \\
		\hline
		T2K & $M(\rho)$ & 0.7 & 0.7 \\
		\hline
		& $N_{l_1}(\rho)$ & 0.6 & 0.6 \\
		\hline
		KamLAND  & $M(\rho)$ & 16.5 & 11  \\
		\hline
		& $N_{l_1}(\rho)$ & 11 & 7\\
		\hline
		JUNO & $M(\rho)$ & 5 & 3.3 \\
		\hline
		& $N_{l_1}(\rho)$ & 3.5 & 2.4 \\
		\hline
		Daya Bay & $M(\rho)$ & $\approx 0$ & $\approx 0$ \\
		\hline
		& $N_{l_1}(\rho)$ & $\approx 0$ & $\approx 0$ \\
		\hline
	\end{tabular}
	\label{tabIII}
\end{table}

From Table \ref{tabIII}, we can see that NSI can enhance the violation in $M(\rho)$ upto 16.5\% with respect to vacuum and 11\%  as compared to the SM interaction in the case of KamLAND, while for JUNO, the corresponding enhancements are only upto  5\% and 3.3\%. For Daya Bay, no significant modification in $M(\rho)$ is visible  in the presence of NSI. In case of DUNE, NSI can increase $M(\rho)$ violation upto 4.3\% while it is upto 2.4\% for MINOS. Having shorter baseline, results for T2K are similar to that of Daya Bay experiment. KamLAND can also provide most substantial effect in $N_{l_1}(\rho)$ in the presence of NSI which can amplify NAQC violation upto 11\% and 7\%  as compared to vacuum and SM interaction, respectively. For JUNO,  the corresponding increments are restricted to  3.5\% and 2.4\% whereas for DUNE these enhancements are upto 4\% and 2.3\%. The results for MINOS are almost the same as obtained for DUNE experiment. Further, owing  to shorter baseline, NSI can produce no noteworthy change in $N_{l_1}(\rho)$ for Daya Bay and T2K experiments. Hence, it can be inferred that $N_{l_1}(\rho)$ parameter, being a comparatively stronger witness of nonclassicality, shows lesser sensitivity to the NSI effects in comparison to the Bell parameter $M(\rho)$.

In Fig. (\ref{fig3}), we have plotted factors $f_a$, $f_b$ and $f_c$, given in Eq. \eqref{r2}) for DUNE (upper left), MINOS (upper middle), T2K (upper right), KamLAND (lower left), JUNO (lower middle) and Daya Bay (lower right) experiments. Both $N_{l_1}(\rho)$ and $M(\rho)$ are functionals of these factors as shown in Eq. \eqref{Nl1} and \eqref{mrho}. It can be seen that $f_b$ is the most sensitive to the NSI as well as SM effects, while $f_c$ shows slightly lesser sensitivity to such effects. $f_a$ has the least sensitivity for both NSI and SM matter interaction. Further, these effects are more distinguishable in case of KamLAND and JUNO (slightly lesser). In case of DUNE, NSI effects are slightly separable than the effects of SM.  However, plots representing SM and vacuum scenario overlap showing no discrimination. In case of SBL reactor experiment Daya Bay, all these factors show negligible deviation in the presence of NSI. $M(\rho)$ is a linear combination of all the three factors, while $N_{l_1}(\rho)$ has dependence on $f_b$ only. It can be clearly observed by comparing Fig. (\ref{fig1}), (\ref{fig2}) and (\ref{fig3}) that the nature of both $M(\rho)$ and $N_{l_1}(\rho)$ in all six experimental set-ups are similar to that of $f_b$. This justifies the observed deviation in $M(\rho)$ and $N_{l_1}(\rho)$ in different energy domains.

\section{Conclusion}\label{sec5}
In this work we study the effect of NSI on NAQC parameter $N_{l_1}(\rho)$  as well as on Bell's inequality parameter $M(\rho)$ for two flavour neutrino oscillations in the context of different accelerator and reactor experimental set-ups. 
The parameter $N_{l_1}(\rho)$ is found to show stronger evidence of quantumness as compared to the Bell-parameter $M(\rho)$. We observe that NSI can enhance the violation of NAQC and $M(\rho)$  at higher energies in comparison to the SM and vacuum scenarios. In the case of LBL reactor experiment experiment KamLAND, this distinction is significantly visible in specific energy regions. However, the Bell-parameter $M(\rho)$ seems to be more sensitive to the new physics effects in comparison to the NAQC parameter.

Therefore the measurement of non-local correlations can provide an alternative platform to probe new physics in the neutrino systems. Owing to the larger sensitivity to NSI, the Bell's inequality parameter is more suited to investigate such effects in the neutrino oscillations. Further, as the dynamics of neutrino oscillations is driven by weak interactions, the effects of environmental dissipation will appear at a much smaller level in the neutrino systems as compared to the optical and condensed matter systems. Therefore the neutrino systems lay out an unique opportunity to design experimental set-ups to test the foundational aspects of quantum mechanics as well as to perform various quantum information tasks. As the strongest measures of quantum correlations are shown to be affected by NSI, its effects should be incorporated while performing such tasks.

The current analysis performed under the framework of two flavour neutrino oscillations shows observable effects on two of the strongest measures of quantum correlations. The effects of NSI are expected to be more profound for three flavour neutrino oscillations. Therefore it is worth extending this work within the framework of three flavour oscillations. Such an analysis also enables us to study the sensitivity of these measures to the $CP$ violating phase and the sign of the neutrino mass square difference.

\bibliographystyle{apsrev4-2}




\end{document}